\RequirePackage[l2tabu, orthodox]{nag}
\RequirePackage{snapshot}

\documentclass{article}

\usepackage{srcltx}
\usepackage[dvips,letterpaper]{geometry}

\usepackage{amsmath}
\usepackage{amsfonts}
\usepackage{amssymb}

\usepackage{bm}

\usepackage{subfigure}
\usepackage{epsfig}
\usepackage{color}

\usepackage{fullpage}
\usepackage{setspace}

\usepackage[normalem]{ulem}

\usepackage{apacite}

       \usepackage{fouriernc}

\newtheorem{proposition}{Proposition}

\def\QED{\mbox{$\square$}}
\def\proof{\noindent{\it Proof:~}}
\def\endproof{\hspace*{\fill}~\QED\par\endtrivlist\unskip}

\begin{document}

\title{The McDonald Normal Distribution}

\date{}

\author{Gauss M. Cordeiro$^\ast$
\and
R. J. Cintra$^\ast$
\and
L. C. R\^ego%
\thanks{G. M. Cordeiro, R. J. Cintra and L. C. R\^ego
are with the
Departamento de Estat\'istica,
Universidade Federal de Pernambuco,
e-mail: \{gauss,rjdsc,leandro\}@de.ufpe.br\space.}
\and
E. M. M. Ortega%
\thanks{E. M. M. Ortega
is with the
Departamento de Ci\^encias Exatas,
Universidade de S\~ao Paulo,
e-mail:
edwin@esalq.usp.br\space.}
}

\maketitle

\begin{abstract}
A five-parameter distribution called the McDonald normal
distribution is defined and studied. The new distribution contains, as special
cases, several important distributions discussed in the literature,
such as the normal, skew-normal, exponentiated normal, beta normal and Kumaraswamy normal distributions, among
others.  We obtain its ordinary moments, moment generating
function and mean deviations. We also
derive the ordinary moments of the order statistics. We use the method
of maximum likelihood to fit the new distribution and illustrate its
potentiality with three applications to real data.

\end{abstract}

\section{Introduction}

For an arbitrary parent cumulative distribution function (cdf) $G(x)$,
the probability density function (pdf) $f(x)$ of the new class
of McDonald generalized distributions (denoted with the
prefix ``Mc'' for short) is defined by
\begin{eqnarray}\label{pdf}
f(x)
=
\frac{c}{B\left(a,b\right)}
\,g(x)\,G^{ac-1}(x)\left\{1-G^c(x)\right\}^{b-1},
\end{eqnarray}
where $a>0$, $b>0$ and $c>0$ are additional shape parameters which aim to
introduce skewness and to vary tail weight and $g(x)=dG(x)/\mathrm{d}x$.
The class of distributions (\ref{pdf}) includes as special
sub-models the beta generalized~\shortcite{eugene2002beta}
for $c=1$ and Kumaraswamy (Kw) generalized distributions~\cite{cordeiro2010family}
for $a=1$. If $X$ is a random variable with density (\ref{pdf}),
we write $X\sim \operatorname{Mc-G}(a,b,c)$.
The density function (\ref{pdf})
will be most tractable when $G(x)$ and $g(x)$
have simple analytic expressions. The corresponding cumulative
function
is
\begin{equation}
\label{cdf}
F(x)
=
I_{G(x)^c}(a,b)
=
\frac{1}{\operatorname{B}(a,b)}
\int_0^{G(x)^c}\omega^{a-1}(1-\omega)^{b-1}
\mathrm{d}\omega,
\end{equation}
where $I_x(a,b)=\operatorname{B}(a,b)^{-1}
\int_0^x\omega^{a-1}(1-\omega)^{b-1}\mathrm{d}\omega$ denotes the incomplete beta function ratio~\cite{gradshteyn2000table}.

Equation~(\ref{cdf}) can also be rewritten as follows
\begin{equation}\label{cdf1}
F(x)
=
\frac{G(x)^{ac}}{a\operatorname{B}(a,b)}
\,_2F_{1}
\left(
a, 1-b; a+1; G(x)^c
\right)
,
\end{equation}
where $_2F_{1}\left(a, b; c; x \right)=\operatorname{B}(b,c-b)^{-1}\int_0^1\frac{t^{b-1}(1-t)^{c-b-1}}{(1-tz)^a}\mathrm{d}t$ is the well-known hypergeometric function~\cite{gradshteyn2000table}.

Some mathematical properties of the cdf $F(x)$ for any Mc-G distribution
defined from a parent $G(x)$ in equation (\ref{cdf1}), could, in principle, follow
from the properties of the hypergeometric function, which are well established in the
literature~\cite[Sec.~9.1]{gradshteyn2000table}.

One major benefit of this class is its ability of fitting skewed data that
can not be properly fitted by existing distributions. Application of $X=G^{-c}(V)$ to a
beta random variable $V$ with positive parameters $a$ and $b$ yields $X$ with cumulative
function (\ref{cdf}).

The associated hazard rate function (hrf) is
\begin{align*}
\tau(x)=
\frac{c\,g(x)\,G^{ac-1}(x)
\left\{
1-G^c(x)
\right\}^{b-1}}%
{\operatorname{B}(a,b)\{1-I_{G(x)^c}(a,b)\}}.
\end{align*}
The Mc-G family of densities allows for greater flexibility of its
tails and can be widely applied in many areas of engineering and
biology.

In this note, we introduce and study the McN distribution for which
its density is obtained from (\ref{pdf}) by taking $G(\cdot)$
and $g(\cdot)$ to be the cdf and pdf of the normal $N(\mu,\sigma^2)$
distribution. The McN density function becomes
\begin{align*}
f(x)
=
\frac{c}{\operatorname{B}\left(a,b\right)\sigma}\,
\phi\left(\frac{x-\mu}{\sigma}\right)
\left\{
 \Phi
 \left(
  \frac{x-\mu}{\sigma}
 \right)
\right\}^{ac-1}
\left\{
1-\Phi
 \left(
  \frac{x-\mu}{\sigma}
 \right)^{c}
\right\}^{b-1}
,
\end{align*}
where $x\in \mathbb{R}$, $\mu\in \mathbb{R}$ is a location parameter,
$\sigma>0$ is a scale parameter, $a$, $b>0$ and $c>0$ are shape parameters,
and $\phi(\cdot)$ and $\Phi(\cdot)$ are the pdf and cdf of the standard
normal distribution, respectively. A random variable with density function
$f(x)$ as above is denoted by
$X\sim \operatorname{McN}(a,b,c,\mu,\sigma^2)$.
For $\mu=0$ and $\sigma=1$, we obtain the standard McN distribution.
Further, the McN distribution with $a=2$ and $b=c=1$ reduces to the skew-normal
distribution~\cite{Azzalini1985} with shape parameter equal to one.

The paper is outlined as follows. Section~\ref{section.expansiodensity} provides some expansions
for the density of the McN distribution.
In Section~\ref{section.bimodality},
we analyze the bimodality properties of the McN distribution.
In Section~\ref{section.moments},
we derive two simple expansions for its moments.
In Sections~\ref{section.generating} and \ref{section.meandeviations}, we obtain the
moment generating function (mgf) and mean deviations, respectively.
We derive, in Section~\ref{section.orderstatistics}, an expansion for the density of the order
statistics. Section~\ref{section.propertiesorderstatistics} provides two representations for the moments
of the order statistics and an explicit expression for the mgf.
In Section~\ref{section.hazard}, we derive the hazard rate function and analyze its limiting behavior.
In Section~\ref{section.entropy}, the Shannon entropy is derived.
Some inferential tools are discussed in Section~\ref{sec:est}. Applications to three real data sets
are illustrated in Section~\ref{section.applications}.
Section~\ref{section.conclusions} ends with some conclusions.

\section{Expansion for the Density}
\label{section.expansiodensity}

Some useful expansions for \eqref{pdf} and \eqref{cdf} can be derived using
the concept of exponentiated distributions.
Here and henceforth, for an arbitrary parent cdf $G(x)$,
we define a random variable $Y$ having the exponentiated
G distribution with parameter $a>0$,
say $Y\sim \operatorname{Exp-G}(a)$, if its
cdf and pdf are given by
\begin{align*}
H_a(x)=G^a(x)
\quad
\text{and}
\quad
h_a(x)=a \, g(x) \, G^{a-1}(x),
\end{align*}
respectively.
The properties of exponentiated distributions have
been studied by many authors in recent years.
In particular,
the exponentiated Weibull~\cite{mudholkar1995reanalysis},
exponentiated Pareto~\shortcite{gupta1998lehman},
exponentiated exponential~\shortcite{gupta2001exponentiated},
and
exponentiated gamma~\cite{nadarajah2007drought}
 distributions are well documented.

By expanding the binomial in (\ref{pdf}), we obtain
\begin{align*}
f(x)=\frac{c}{\operatorname{B}(a,b)}\,g(x)\,
\sum_{k=0}^\infty
(-1)^k\binom{b-1}{k}
G^{(k+a)c-1}(x)
\end{align*}
and then
\begin{eqnarray}\label{exppdf}
f(x)=\sum_{k = 0}^\infty w_k\,h_{(k+a)c}(x),
\end{eqnarray}
where $h_{(k+a)c}(x)$ has the $\operatorname{Exp-G}[(k+a)c]$ distribution and the
weights $w_k$ are given by
\begin{align*}
w_k
=
\frac{(-1)^k \, \binom{b-1}{k}}{(k+a)\operatorname{B}(a,b)}
.
\end{align*}
The $\operatorname{Mc-G}$ density function is then a linear combination of exponentiated G
densities. The properties of the $\operatorname{Mc-G}$ distribution can be obtained by
knowing those of the corresponding exponentiated distributions. Integrating
(\ref{exppdf}), we obtain
\begin{eqnarray*}
F(x)=\sum_{k = 0}^\infty
w_k \,H_{(k+a)c}(x)
.
\end{eqnarray*}

From now on, we work with a random variable $Z$ having the standard
$\operatorname{McN}(a,b,c,0,1)$
distribution.
The density of $Z$ reduces to
\begin{eqnarray*}
f(x)=\frac{c}{\operatorname{B}(a,b)}\,\phi(x)\Phi(x)^{ac-1}\left\{1-\Phi(x)^{c}\right\}^{b-1}.
\end{eqnarray*}

\begin{figure}

\begin{center}
\subfigure[]{\epsfig{file=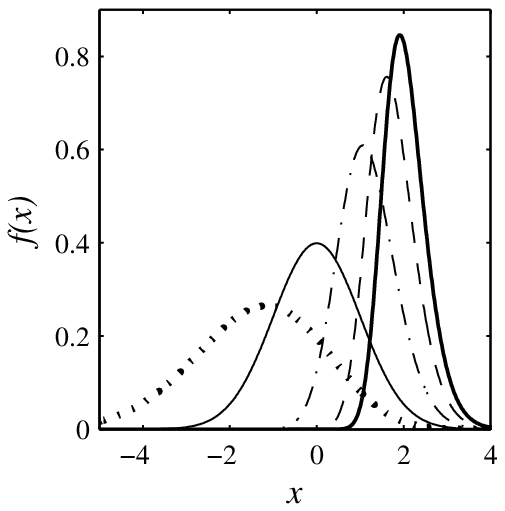}}
\subfigure[]{\epsfig{file=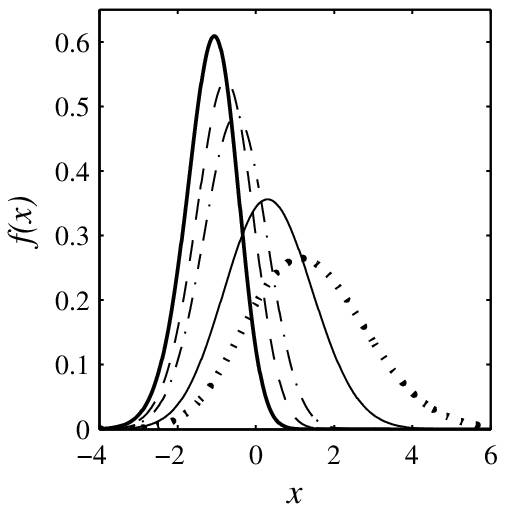}}
\subfigure[]{\epsfig{file=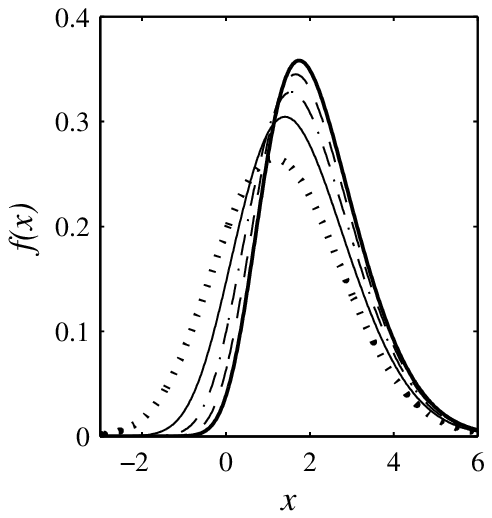}}
\subfigure[]{\epsfig{file=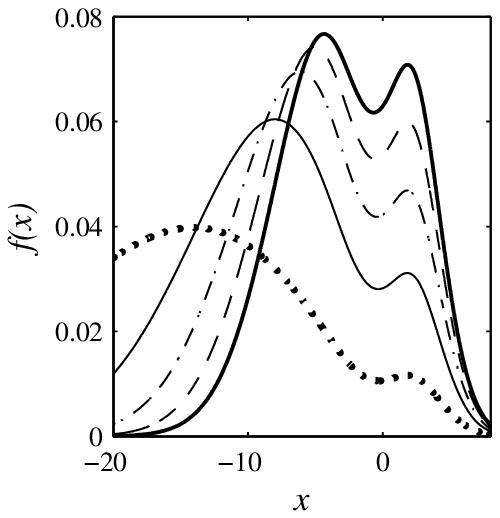}}
\subfigure[]{\epsfig{file=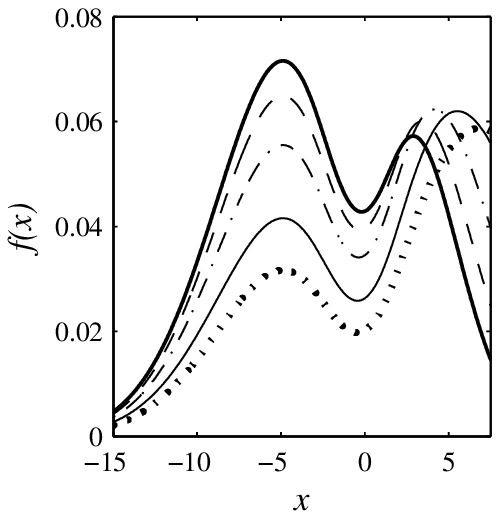}}
\subfigure[]{\epsfig{file=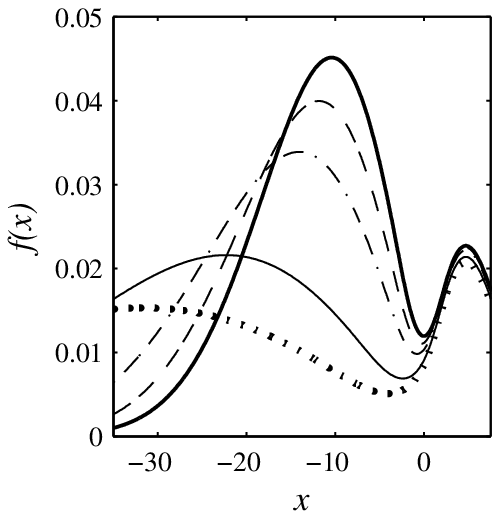}}
\caption{Plots of the $\operatorname{McN}(a,b,c,0,1)$ density for some parameter values.
(a) $a\in\{0.3,1,5,15,30\}$,  $b=1$, and $c=1$,
(b) $a=1$, $b\in\{0.3, 0.7, 2, 3, 5\}$, and $c=1$,
(c) $a=1$, $b=0.3$, and $c\in\{1, 2, 3, 4, 5\}$,
(d) $a = \{0.01, 0.03, 0.05, 0.07, 0.09 \}$, $b=0.15$, and $c=0.5$,
(e) $a = 0.05$, $b = \{ 0.02, 0.03, 0.05, 0.07, 0.09 \}$, and $c=0.75$,
(f) $a=0.01$, $b=0.04$, and $c=\{ 0.1, 0.2, 0.5, 0.7, 0.9 \}$.
In all cases,
plots are dotted, solid, dot-dashed, dashed, and bold solid, respectively.}
\label{fig.fdp}
\end{center}
\end{figure}

Plots of the McN density for selected parameter values are given in
Figure~\ref{fig.fdp}.

Using~\eqref{exppdf}, we can write
\begin{eqnarray}\label{exppdf1}
f(x)=\sum_{k = 0}^\infty t_k\,\Phi(x)^{(k+a)c-1}\phi(x),
\end{eqnarray}
where
$$t_k = t_k(a,b,c)=\frac{(-1)^k c\, \binom{b-1}{k}}{B\left(a,b\right)}.$$

We can obtain an expansion for $\Phi(x)^{\beta}$ for $\beta>0$
real non-integer given by
\begin{eqnarray*}
\Phi(x)^{\beta}=\sum_{j=0}^\infty \sum_{r=0}^j (-1)^{j+r}
\binom{\beta}{j}\binom{j}{r} \Phi(x)^r.
\end{eqnarray*}
We can substitute $\sum_{j=0}^\infty\sum_{r=0}^j$ for
$\sum_{r=0}^\infty\sum_{j=r}^\infty$ to obtain
\begin{equation}\label{powerG}
\Phi(x)^{\beta}=\sum_{r=0}^\infty s_r(\beta) \, \Phi(x)^r,
\end{equation}
where
\begin{align*}
s_r(\beta)=\sum_{j=r}^\infty (-1)^{r+j}
\binom{\beta}{j}\binom{j}{r}
.
\end{align*}
Combining
(\ref{exppdf1}) and (\ref{powerG}),
the McN density function can be expressed as
\begin{eqnarray}\label{exppdf2}
f(x)=\sum_{k,r=0}^\infty t_k\,s_r((k+a)c-1)\,\Phi(x)^r\phi(x).
\end{eqnarray}
Expansions (\ref{exppdf}), (\ref{exppdf1}), and (\ref{exppdf2}) are the main
results of this section.

\section{Bimodality}

\label{section.bimodality}

The analysis of the critical points of the McN density function furnishes
a natural path for characterizing the distribution shape and
quantifying the number of modes.
Taking the normalization
$z=\frac{x-\mu}{\sigma}$,
we have
%the first derivative of $f(z)$ with respect
%to $z$ reduces to
\begin{align*}
\frac{\partial f(z)}{\partial z}=
&
%\frac{\Gamma(a+b)}{\Gamma(a)\Gamma(b)}
\frac{c}{\mathrm{B}(a,b)}
\phi(z)
[\Phi(z)]^{ac-2}
[1-\Phi^c(z)]^{b-2}\\
&
\times
\Big\{
(ac-1)\phi(z)[1-\Phi^c(z)]
-
c(b-1)\phi(z)\Phi^c(z)
-
z \Phi(z) [1-\Phi^c(z)]
\Big\}.
\end{align*}
We refer to the term in curly brackets as $s(z)$.
At the critical points, where $\frac{\partial f(z)}{\partial z}=0$,
we have $s(z)=0$,
since the remaining terms of $\frac{\partial f(z)}{\partial z}$ are
strictly positive.
Hence, the critical points satisfy the following implicit equation
\begin{align}
\label{criticalpoints}
z
=
[1-c(a+b-1)]
%(c+1-ac-bc)
\phi(z)\frac{\Phi^{c-1}(z)}{1-\Phi^c(z)}
+
(a-1)
\frac{\phi(z)}{\Phi(z)[1-\Phi^c(z)]}
.
\end{align}
By analyzing this expression, the bimodality conditions for the McN density can be established.

%\begin{figure}
%\centering
%\subfigure[]{\epsfig{file=figure1a.eps}}
%\qquad
%\subfigure[]{\epsfig{file=figure1b.eps}}
%\caption{
%(a)~Implict curve for the critical values of $f(z)$ for a fixed value of $b=0.15$ and several values of $c\in\{0.5, 0.75, 1, 1.25, 1.5\}$.
%(b)~Associated density plots for $a=0.1$ (dashed) and $a=0.2$ (dot-dashed).
%}
%\label{fig1}
%\end{figure}

%In~\cite{famoye2004bimodality}, it is claimed that the solutions
%of (\ref{criticalpoints}) are the modes of the distribution.
%In fact, there are solutions of this equation that can be
%local minimum of the density function, thus they are not modes. For example,

\begin{figure}
\centering
\subfigure[]{\epsfig{file=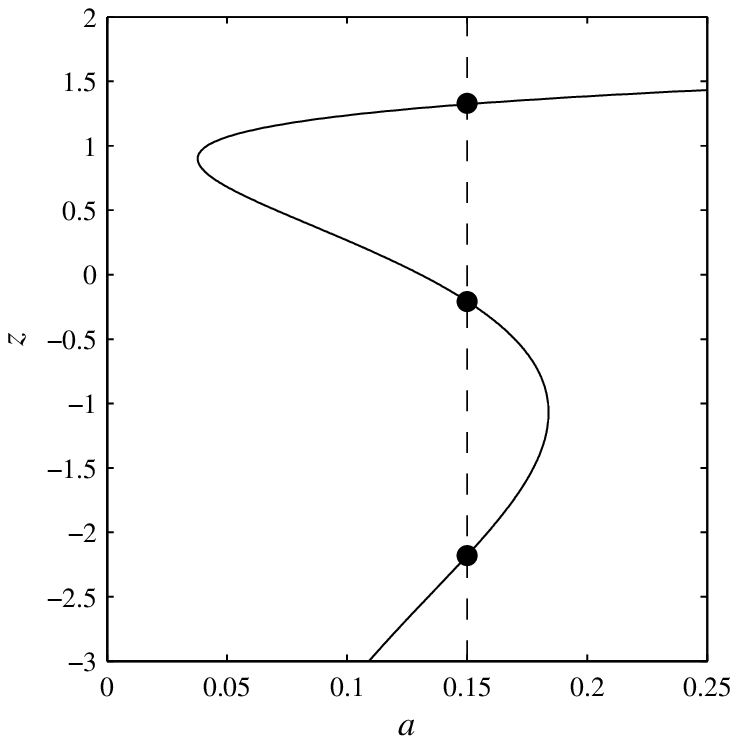}}
\quad
\subfigure[]{\epsfig{file=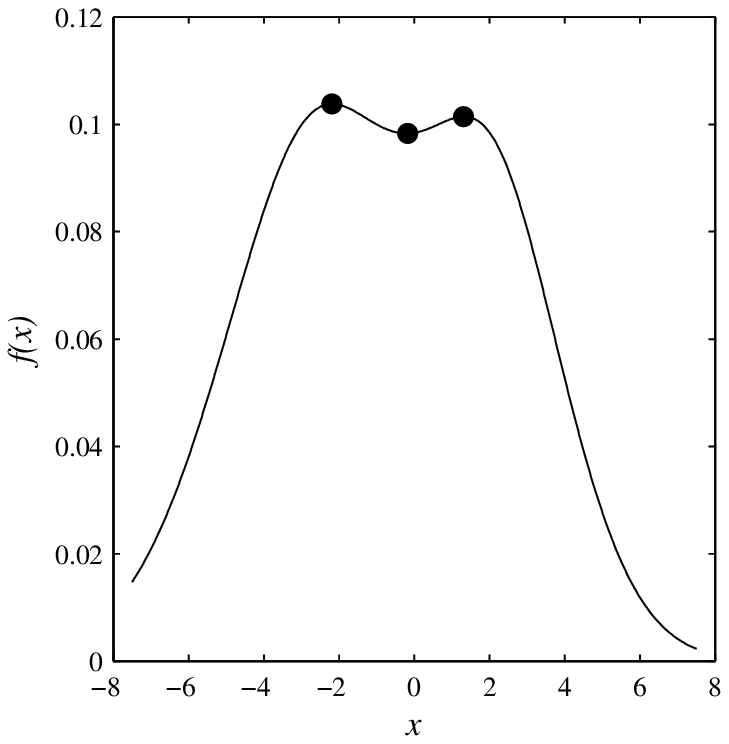}}
\caption{%
(a) $b=0.18$, $c=0.75$;
(b) $a=0.15$, $b=0.18$, $c=0.75$}
\label{fig2}
\end{figure}

As a particular example,
Figure~\ref{fig2}(a) gives the plot of the solutions of (\ref{criticalpoints})
in terms of $a$ for a fixed value of $b=0.18$.
For $a=0.15$, there are three solutions indicated by filled dots ($\bullet$).
%However, for $\alpha=0.2$, only one critical point is found ($\circ$).
%
Figure~\ref{fig2}(b) provides the corresponding density plots.
For $a=0.15$,
only two of the marked points are indeed modes of the density function,
since the remaining point characterizes a local minimum.

\begin{figure}
\centering
\subfigure[]{\epsfig{file=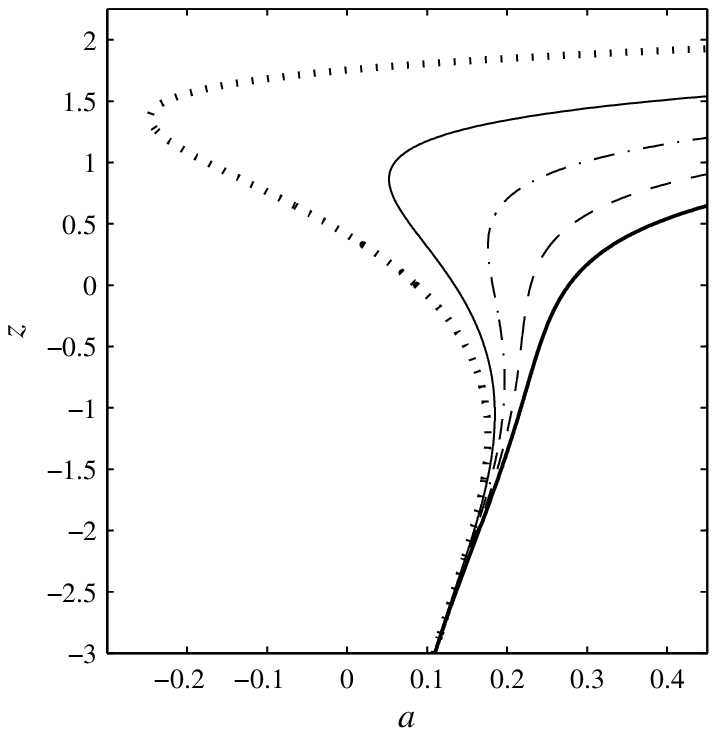}}
\quad
\subfigure[]{\epsfig{file=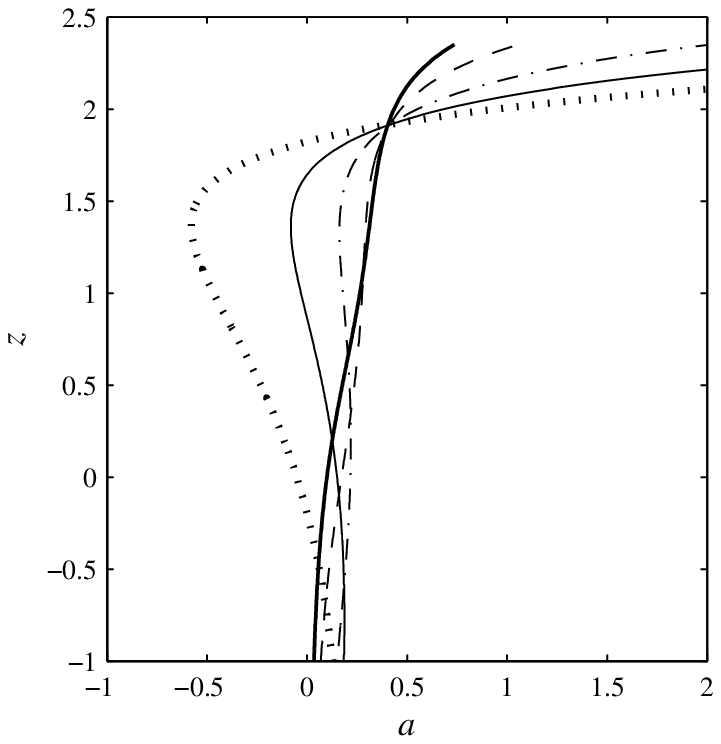}}
\quad
\subfigure[]{\epsfig{file=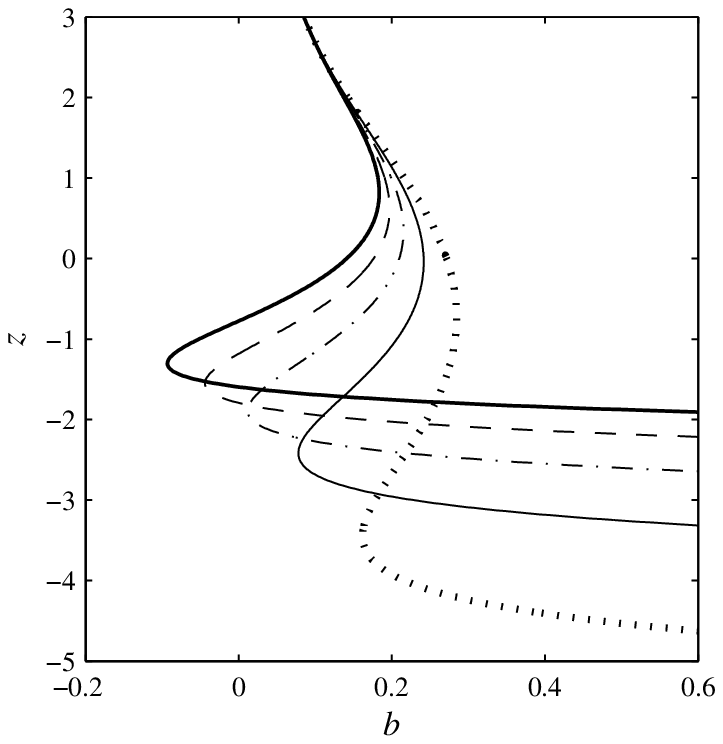}}
\quad
\subfigure[]{\epsfig{file=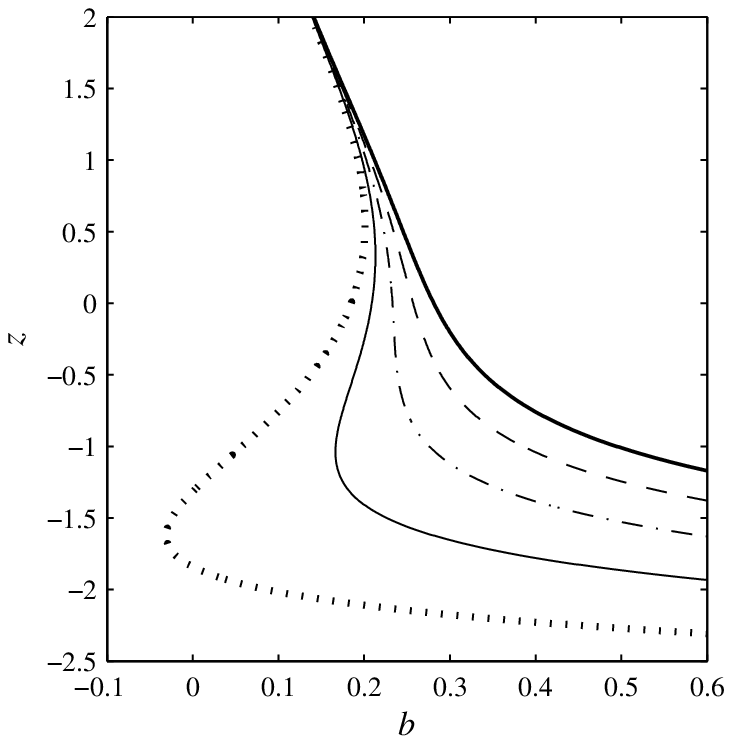}}

\caption{%
(a) $c = 0.75$, $b\in\{0.15, 0.1825, 0.215, 0.2475, 0.28\}$,
(b) $b = 0.15$, $c\in\{0.5, 1, 2, 5, 10\}$,
(c) $c = 0.8$, $a\in\{0.15, 0.1825, 0.215, 0.2475, 0.28 \}$,
(d) $a = 0.15$, $c\in\{0.25, 0.45, 0.65, 0.85, 1.05 \}$.
}
\label{fig3}
\end{figure}

Figure~\ref{fig3}(a)-(b) give
plots of the solutions of (\ref{criticalpoints})
in terms of $a$ for a fixed value of $c$ and varying values of $b$;
and for a fixed value of $b$ and varying values of $c$,
respectively.
%For $a=0.1$, there are three solutions indicated by filled dots ($\bullet$).
%However, for $a=0.2$, only one critical point is found ($\circ$).
%
Analogously,
Figure~\ref{fig3}(c)-(d) provide plots
in terms of $b$.
%
%For given $b$, the choice of $a$ determines the number of modes of $f(z)$.
Additionally, we note that the only parts of the implicit curve
with probabilistic meaning are those situated in the region $a>0$ and $b>0$.

In order to determine which critical points are modes of the distribution,
we should consider the sign of the second derivative at the critical points.
In particular, a mode of $f(z)$ is a critical point with non-positive second derivative.
At the critical points, we have
\begin{align*}
\frac{\partial^2 f(z)}{\partial z^2}=&
%\frac{\Gamma(a+b)}{\Gamma(a)\Gamma(b)}
\frac{c}{\mathrm{B}(a,b)}\,
\phi(z)\,[\Phi(z)]^{ac-2}\,[1-\Phi^c(z)]^{b-2}\,
\frac{\partial s(z)}{\partial z}.
\end{align*}
The sign of $\frac{\partial^2 f(z)}{\partial z^2}$ is the
same of $\frac{\partial s(z)}{\partial z}$.
Then, at a mode, the condition $\frac{\partial s(z)}{\partial z}\leq 0$ holds.
Explicit evaluation of $\frac{\partial s(z)}{\partial z}$ yields
\begin{align}
\label{derivativeofs(z)}
\frac{\partial s(z)}{\partial z}
=
-\Phi(z)\,[1-\Phi^c(z)]
-
c z\,\phi(z)
\left[a-(a+b)\,\Phi^c(z)\right]
+
c
[1-c(a+b-1)]
\phi^2(z)\Phi^{c-1}(z).
\end{align}
We are now able to prove the following result.
\begin{proposition}
If $c(1-b)=(2^c-1)(1-ac)$ and $a>\frac{\pi}{2c^2}(2^{-c}-1)+\frac{1}{c}$,
then $z=0$ is a modal point of the McN distribution.
\end{proposition}

\proof
If $s(0)=0$ and $\frac{\partial s(0)}{\partial z}<0$,
then $z=0$ is a modal point of the McN distribution.
From the definition of $s(z)$,
it follows that $s(0)=0$ if and only if $c(1-b)=(2^c-1)(1-ac)$.
Letting $z=0$ in (\ref{derivativeofs(z)}) yields:
\begin{align*}
\frac{\partial s(0)}{\partial z}
=
-
\frac{1}{2}
\left(
1-\frac{1}{2^c}
\right)
+
\frac{1}{2\pi}\frac{1}{2^{c-1}}
c
[1-c(a+b-1)]
.
\end{align*}
Using the first order condition $c(1-b)=(2^c-1)(1-a)$ and imposing
$\frac{\partial s(0)}{\partial z}<0$
gives
$a>\frac{\pi}{2c^2}(2^{-c}-1)+\frac{1}{c}$.
\endproof

We consider the variational behavior of the critical points of $f(z)$ with respect
to changes in the parameter $a$. From equation (\ref{criticalpoints}), the first derivative
of $z$ with respect to $a$ is
\begin{align*}
\frac{\partial z}{\partial a}
=
&
% \frac{\phi(z_0)(1-\Phi(z_0))}%
% {\Phi(z_0)(1-\Phi(z_0)) + z_0\phi(z_0)(1-2\Phi(z_0)) - \phi^2(z_0)(2-a-b) + z_0\phi(z_0) \left[(2-a-b) \Phi(z_0) + (a-1) \right]}
% \\
% =
% &
\frac{c \phi(z)\,[1-\Phi^c(z)]}{\Phi(z)\,[1-\Phi^c(z)]+
c z
\phi(z)
\left[a-(a + b)\,\Phi^c(z)\right]
-c[1-c(a+b-1)]
\phi^2(z)\Phi^{c-1}(z)}.
\end{align*}

Since $\phi(z)[1-\Phi^c(z)]>0$, the sign of $\frac{\partial z}{\partial a}$
depends entirely on the behavior of the denominator term.
Moreover, except for the sign,
this denominator is equal to $\frac{\partial s(z)}{\partial z}$.
Thus, for any $z$, we have
\begin{align*}
\frac{\partial z}{\partial a}
=
\frac{c\phi(z)[1-\Phi^c(z)]}%
{-\frac{\partial s(z)}{\partial z}}
\quad
\text{and}
\quad
\operatorname{sign}\left(\frac{\partial z}{\partial a}\right)
=
-\operatorname{sign}\left(\frac{\partial s(z)}{\partial z}\right),
\end{align*}
%
%Therefore, for any $z$, we obtain
%\begin{align*}
%\mathrm{sign}\left(\frac{\partial z}{\partial a}\right)
%&=
%-\mathrm{sign}\left(\frac{\partial s(z)}{\partial z}\right),
%\end{align*}
%
where $\operatorname{sign}(\cdot)$ is the sign function.
Further,
if $\frac{\partial s(z)}{\partial z}$ is negative
at a critical point $z$,
then $z$ must be a mode which is an increasing function of $a$.
%
%
%However,
%nothing prevents the existence of a mode for which
%$\frac{\partial s(z)}{\partial z}$ vanishes.
%Indeed, in the case $a=b =1-\pi/4$, it can be shown that
%\begin{align*}
%\frac{\partial f(0)}{\partial z}
%=
%\frac{\partial^2 f(0)}{\partial z^2}
%=
%\frac{\partial^3 f(0)}{\partial z^3}
%=
%0
%\quad
%\text{and}
%\quad
%\frac{\partial^4 f(0)}{\partial z^4}
%&=
%\frac{4\sqrt{2}}{\pi}
%\frac{\Gamma(3/2-\pi/4)}{\Gamma(1-\pi/4)}
%\left(\frac{3}{\pi}-1 \right)<0.
%\end{align*}
%
%Thus, $z=0$ is a mode of the distribution.
%Numerical computations give evidence that
%this is the unique case for which the second derivative vanishes
%at the mode.
%In this situation,
%$\frac{\partial z}{\partial a}$ is undefined.
Nevertheless, it is still true that $z$ is an
increasing function of $a$.
%, as shown in Figure~\ref{fig.inflexao}.
We then state the following proposition.

%\begin{figure}
%\centering
%%\epsfig{file=curva_inflexao.eps}
%%\epsfig{file=figure2.eps}
%\caption{At $a=b=1-\pi/4$, the quantity $\partial z/\partial a$ is undefined.}
%\label{fig.inflexao}
%\end{figure}

\begin{proposition}
If $z$ is a mode location,
then $z$ is an increasing function of $a$.
%i.e.
%$\frac{\partial z}{\partial a}>0$.
\end{proposition}

We now consider the variational behavior of the critical points of $f(z)$ with respect
to changes in $b$. From equation~(\ref{criticalpoints}), the first derivative
of $z$ with respect to $b$ is
\begin{align*}
\frac{\partial z}{\partial b}
=
&
% \frac{\phi(z_0)(1-\Phi(z_0))}%
% {\Phi(z_0)(1-\Phi(z_0)) + z_0\phi(z_0)(1-2\Phi(z_0)) - \phi^2(z_0)(2-a-b) + z_0\phi(z_0) \left[(2-a-b) \Phi(z_0) + (a-1) \right]}
% \\
% =
% &
\frac{c\phi(z)\,\Phi^c(z)}{-\Phi(z)\,[1-\Phi^c(z)]-
c z
\phi(z)
\left[a-(a + b) \Phi^c(z)\right]
+c[1-c(a+b-1)]\,\phi^2(z)\Phi^{c-1}(z)}.
\end{align*}

Since $c\phi(z)\Phi^c(z)>0$, the sign of $\frac{\partial z}{\partial b}$
depends entirely on the behavior of the denominator term.
Moreover, this denominator is equal to $\frac{\partial s(z)}{\partial z}$.
Thus, for any $z$, we have
\begin{align*}
\frac{\partial z}{\partial b}&=\frac{c\phi(z)\Phi^c(z)}%
{\frac{\partial s(z)}{\partial z}}
\quad
\text{and}
\quad
\mathrm{sign}\left(\frac{\partial z}{\partial b}\right)
=
\mathrm{sign}\left(\frac{\partial s(z)}{\partial z}\right).
\end{align*}

\subsection{Modality Regions}

\begin{figure}
\centering
\subfigure[]{\epsfig{file=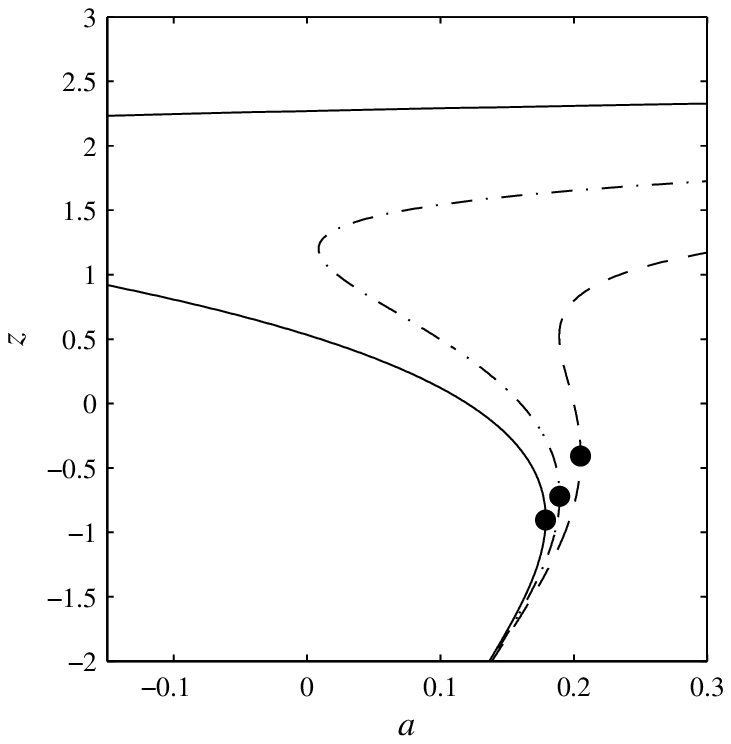}}
\quad
\subfigure[]{\epsfig{file=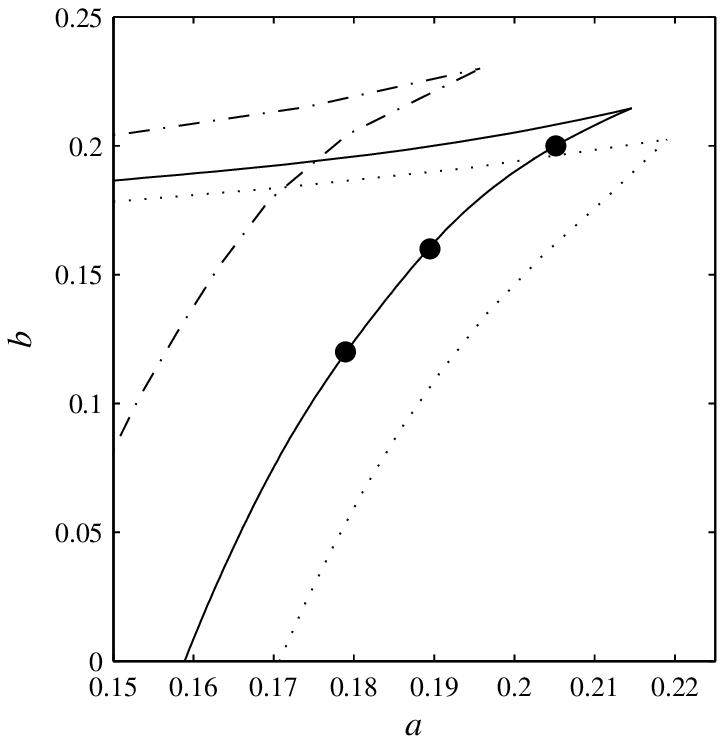}}
\caption{%
(a) $c = 1$, $b \in\{0.12, 0.16, 0.2 \}$
(solid, dash-dotted, dashed),
(b) $c \in \{0.8, 1,  1.5 \}$
(dotted, solid, dash-dotted).
}
\label{fig4}
\end{figure}

From equation (\ref{criticalpoints}),
we can express $a$ in terms of $z$ for fixed values
of $b$ and $c$.
Let $a_{b,c}(z)$ denote this function of $z$ by fixing $b$ and $c$.
Hence,
\begin{align*}
a_{b,c}(z)=
%\frac{z\Phi(z)(1-\Phi(z))-\phi(z)[(2-b)\Phi(z)-1]}{\phi(z)(1-\Phi(z))}
\frac{1}{c}
\frac{z\Phi(z)}{\phi(z)}
-
\frac{(1-b+1/c)\Phi^c(z)-1/c}{1-\Phi^c(z)}.
\end{align*}

%Moreover, for fixed values of $b\in (0,1-\pi/4)$ and $c\in ?$,
Let $a^*_{b,c}$ denote the local maximum of $a_{b,c}(z)$.
A similar analysis of the variation of the critical points
in terms of $b$ for several values of $a$ and $c$ can be made.

Due to its symmetric behavior,
the discussion follows \emph{mutatis mutandis}.
So,
from equation (\ref{criticalpoints}),
we can express $b$ in terms of $z$ for fixed
values of $a$ and $c$.
Let $b_{a,c}(z)$ denote such function given by
\begin{align*}
b_{a,c}(z)=
\frac{a-1/c}{\Phi^c(z)}
- \frac{z[1-\Phi^c(z)]}{c\phi(z)\Phi^{c-1}(z)}
+
%\frac{c+1-a}{c}
1-a+1/c
.
\end{align*}
Let $b^*_{a,c}$ denote the local maximum of $b_{a,c}(z)$.

%Table~\ref{table1} lists the numerical values of
%$b^*_\gamma=a^*_\gamma$
%for several values of $\gamma\in(0,1-\pi/4)$.
%Thus, the region of the parameters $a$ and $b$ for which
%the distribution is bimodal reduces to
%\begin{align*}
%\left\{
%(a,b):
%0<a<\min\left(1-\frac{\pi}{4},a^*_b\right),
%0<b<\min\left(1-\frac{\pi}{4},b^*_a\right)
%\right\}
%.
%\end{align*}
%%
%The curves
%$\{(a,b^*_a):0<a<1-\pi/4\}$
%and
%$\{(a^*_b,b):0<b<1-\pi/4\}$
%which delimit the above region are
%symmetric with respect to the line $a=b$.

For each implicit curve
$a_{b,c}(z)$,
a critical point could be identified and
marked with a bullet.
The abscissa values of these critical points
indicate
the boundary value of $a$
that makes the McN density function
switch behavior from
a bimodal distribution to a unimodal one.
Figure~\ref{fig4}(b)
shows the modality regions of the McN distribution.

\section{Moments}
\label{section.moments}

The moments of $X$ having
the $\operatorname{McN}(a,b,\mu,\sigma)$ distribution are immediately
obtained from the moments of $Z$ following
the $\operatorname{McN}(a,b,0,1)$ distribution
by
$\operatorname{E}(X^n)=E[(\mu+\sigma Z)^n]=\sum_{r=0}^{n}\binom{n}{r}\mu^{n-r}\sigma^r \operatorname{E}(Z^r)$.
So, we can work with the standard McN distribution. We give two representations
for the $n$th moment of the standard McN distribution,
say $\mu_n^{\prime}=\operatorname{E}(Z^n)$.
First, $\mu_n^{\prime}$ can be derived from (\ref{exppdf1}) as
\begin{align*}
\mu_n^{\prime}
=
\sum_{k = 0}^\infty
t_k\,
\int_{-\infty}^{\infty}
x^n\,\Phi(x)^{(k+a)c-1}
\,\phi(x)\mathrm{d}x.
\end{align*}
Setting $u=\Phi(x)$, we can write $\mu_n^{\prime}$ in terms of the standard normal quantile
function $Q(u)=\Phi^{-1}(u)$ as
\begin{eqnarray}\label{moment1}
\mu_n^{\prime}=\sum_{k = 0}^\infty t_k\,\int_{0}^{1} Q(u)^n \, u^{(k+a)c-1}
\mathrm{d}u.
\end{eqnarray}

The standard
normal quantile function can be expanded as~\cite{Steinbrecher2002}
\begin{eqnarray}\label{generalform}
Q(u)=\sum_{k=0}^{\infty}b_k \, v^{2k+1},
\end{eqnarray}
where $v=\sqrt{2\pi}(u-1/2)$ and the $b_k'$s are calculated recursively from
\begin{align*}
b_{k+1}=
\frac{1}{2(2k+3)}
\sum_{r=0}^{k}
\frac{(2r+1)(2k-2r+1)
\,
b_r
\,
b_{k-r}}{(r+1)(2r+1)}
.
\end{align*}
Here, $b_0=1$, $b_1=1/6$, $b_2=7/120$, $b_3=127/7560,\ldots$

By application of an equation in
\citeauthor{gradshteyn2000table}~\citeyear[Sec.~0.314]{gradshteyn2000table}
for a power series raised to a positive integer, $n$, we obtain
\begin{eqnarray}\label{expnint}
Q(u)^n
=
\left(
\sum_{i=0}^\infty a_i \, v^i
\right)^n
=
\sum_{i=0}^\infty
c_{n,i} \, v^i.
\end{eqnarray}
Here, the coefficients $c_{n,i}$ for $i=1,2,\ldots$ are easily
obtained from the recurrence equation
\begin{equation}\label{formulae_c}
c_{n,i}=(i a_0)^{-1}\sum_{m=1}^{i}[m(n+1)-i]\,a_m\,c_{n,i-m},
\end{equation}
where $c_{n,0}=a_0^n$.
The coefficient $c_{n,i}$ can be determined from
$c_{n,0},\ldots,c_{n,i-1}$ and hence from the quantities
$a_0,\ldots,a_{i}$.
Equations (\ref{expnint}) and (\ref{formulae_c})
are used throughout this article. The coefficient $c_{n,i}$ can be given explicitly
in terms of the coefficients $a_i$, although it is not necessary for programming numerically
our expansions in any algebraic or numerical software.

The coefficients $a_i$ in (\ref{expnint}) are defined from those
in (\ref{generalform}) by: $a_i=0$ for $i=0,2,4,\ldots$ and $a_i=b_{(i-1)/2}$
for $i=1,3,5,\ldots$, and then the quantities $c_{n,i}$ can be calculated
numerically from the $a_i'$s by (\ref{formulae_c}). We can easily obtain
from (\ref{moment1})
\begin{eqnarray}\label{moment2}
\mu_n^{\prime}=\sum_{k,i=0}^\infty (2\pi)^{i/2} t_k\,c_{n,i}\,\sum_{r=0}^{i}\frac{\binom{i}{r}(-1)^r 2^{-r}}{(k+a)c+i-r}.
\end{eqnarray}
The moments of the McN distribution can be determined from equation (\ref{moment2}), where
the quantities $c_{n,i}$ are derived from (\ref{formulae_c})
using the $a_i'$s above.

We now provide a second representation for $\mu_n^{\prime}$. The standard
normal cdf can be expressed as
$$\Phi(x)=\frac{1}{2}\left\{1+ \rm{erf}\left(\frac{x}{\sqrt{2}}\right)\right\},\quad x \in \mathbb{R}.$$

The $(n,r)$th probability weighted moment (PWM) (for $n$ and $r$ integers)
of the standard normal distribution is defined by
$$\tau_{n,r}=\int_{-\infty}^{\infty} x^n \Phi(x)^r \phi(x) \mathrm{d}x.$$
By making use of the binomial expansion and interchanging terms,
we have
$$\tau_{n,r}= \frac{1}{2^r\sqrt{2\pi}}\sum_{p=0}^r\binom{r}{p}\int_{-\infty}^{\infty} x^n
\operatorname{erf}\left(\frac{x}{\sqrt 2}\right)^{r-p}\,\exp\left(-\frac{x^2}{2}\right)\mathrm{d}x.$$

Using the series expansion for the error function $\operatorname{erf}(\cdot)$
\begin{align*}
\operatorname{erf}(x)
=
\frac{2}{\sqrt{\pi}}
\sum_{m=0}^\infty
\frac{(-1)^m \, x^{2m+1}}{(2m+1)m!},
\end{align*}
we can determine $\tau_{n,r}$ from equations
(9)-(11)
given by \citeauthor{nadarajah2008order}~\citeyear{nadarajah2008order}.
For $n+r-p$ even, we have
\begin{eqnarray}\nonumber
\tau_{n,r}&=& 2^{n/2}\pi^{-(r+1/2)}
\sum_{\stackrel{p=0} {(n+r-p)\,\text{even}}}^r \binom{r}{p} 2^{-p}
\pi^p\Gamma\left(\frac{n+r-p+1}{2}\right)\times\\
&&
F_A^{(r-p)}
\left(
  \frac{n+r-p+1}{2};\frac{1}{2},\ldots,\frac{1}{2};\frac{3}{2},\ldots,\frac{3}{2};-1,\ldots,-1
\right),\label{pwmnormal}
\end{eqnarray}
where the terms in $\tau_{n,r}$ vanish when $n+r-p$ is odd. The moments of the standard
McN distribution is calculated from equation (\ref{exppdf2}) as
\begin{eqnarray}\label{moment3}
\mu_n^{\prime}=\sum_{k,r=0}^\infty t_k\,s_r((k+a)c-1)\,\tau_{n,r},
\end{eqnarray}
where $\tau_{n,r}$ is given by (\ref{pwmnormal}).
Equations (\ref{moment2})
and (\ref{moment3}) are the main results of this section.

\begin{figure}
\begin{center}
\subfigure[]{\epsfig{file=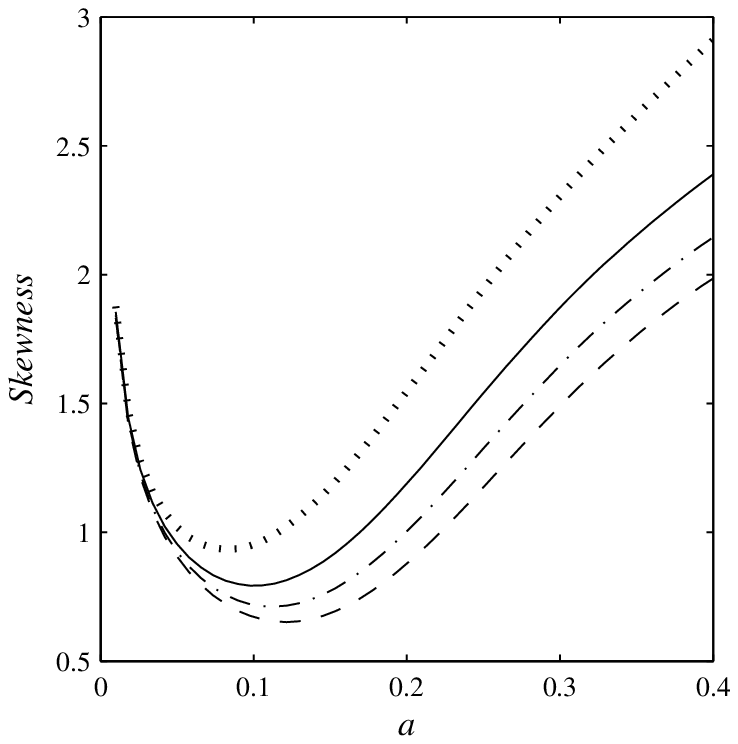}}
\subfigure[]{\epsfig{file=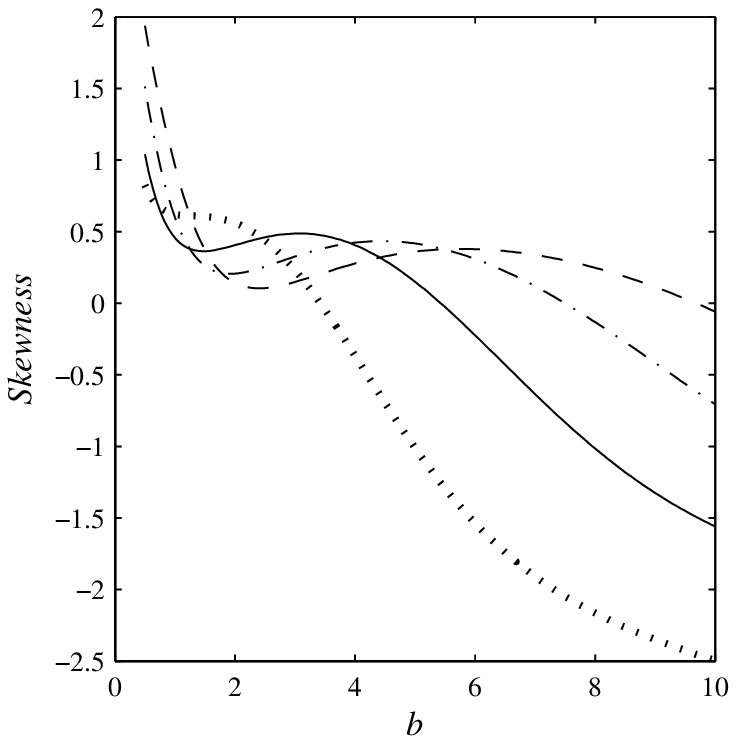}}
\subfigure[]{\epsfig{file=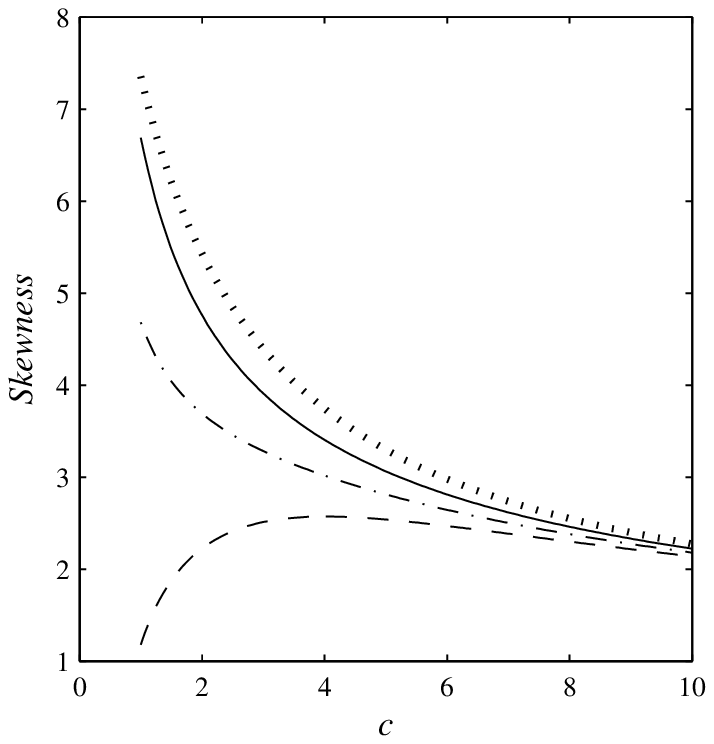}}
\caption{Skewness of the McN distribution.
(a) Function of $a$ for some values of $b\in\{ 0.5, 1, 1.5, 2 \}$ (dotted, solid, dot-dashed, dashed)
with $\mu=0$, $\sigma=1$ and $c=15$.
(b) Function of $b$ for some values of $a\in\{ 0.5, 1, 1.5, 2 \}$ (dotted, solid, dot-dashed, dashed)
with $\mu=-1$, $\sigma=1$ and $c=7$
(c) Function of $c$ for some values of $b\in\{ 0.5, 1, 1.5, 2 \}$ (dotted, solid, dot-dashed, dashed)
with $\mu=0$, $\sigma=1$ and $a=0.01$.
\label{assimetria}}
\end{center}
\end{figure}

\begin{figure}
\begin{center}
\subfigure[]{\epsfig{file=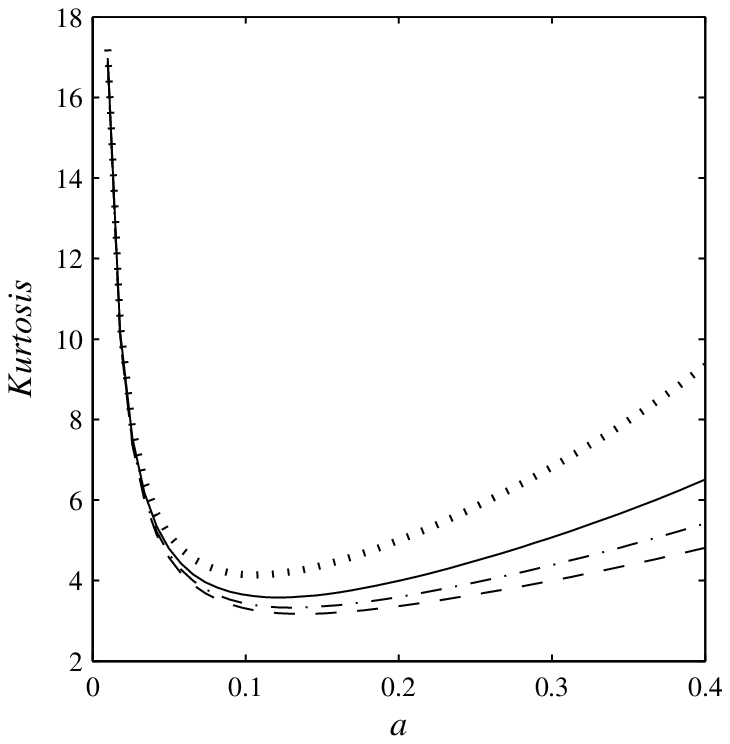}}
\subfigure[]{\epsfig{file=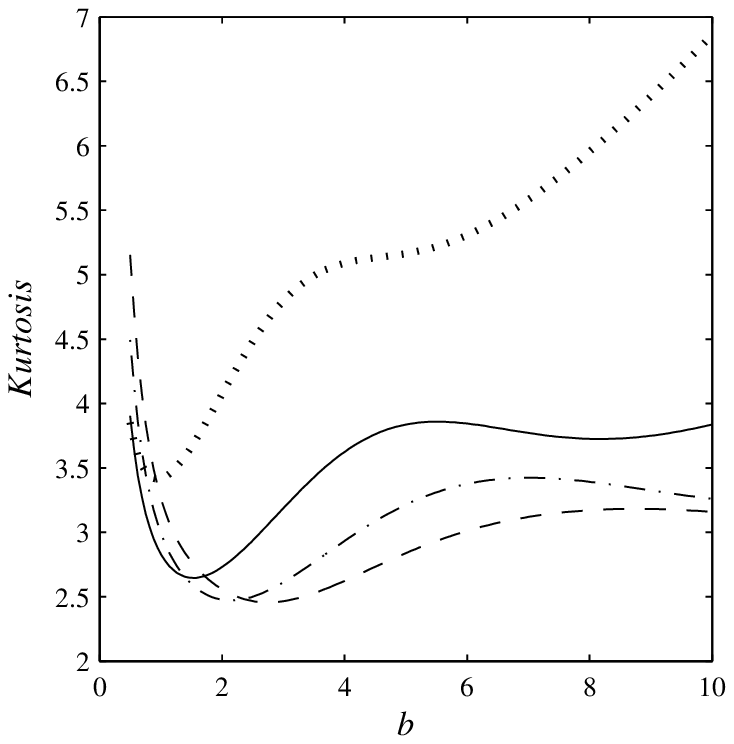}}
\subfigure[]{\epsfig{file=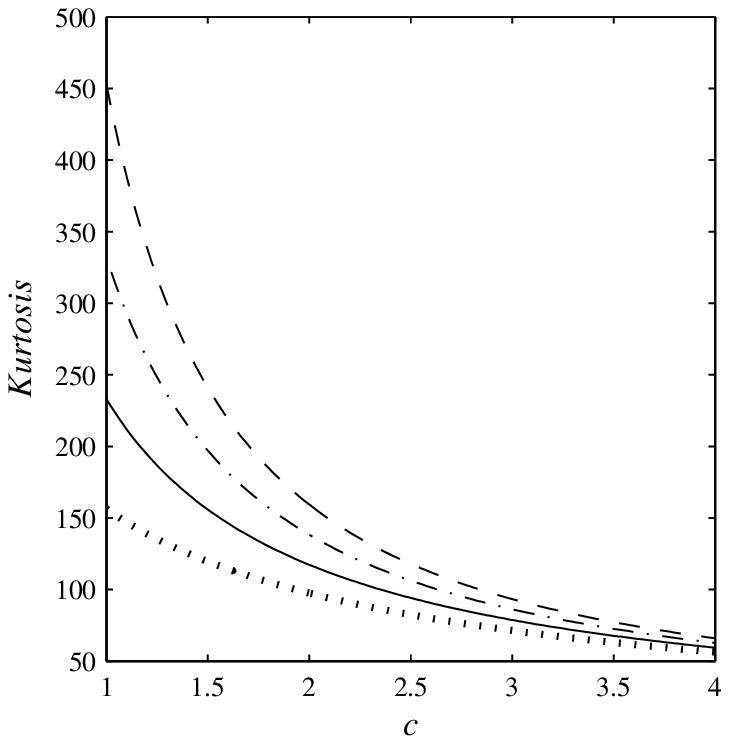}}
\caption{Kurtosis of the McN distribution.
(a) Function of $a$ for some values of $b\in\{ 0.5, 1, 1.5, 2 \}$ (dotted, solid, dot-dashed, dashed)
with $\mu=0$, $\sigma=1$ and $c=15$.
(b) Function of $b$ for some values of $a\in\{ 0.5, 1, 1.5, 2 \}$ (dotted, solid, dot-dashed, dashed)
with $\mu=-1$, $\sigma=1$ and $c=7$
(c) Function of $c$ for some values of $b\in\{ 0.5, 1, 1.5, 2 \}$ (dotted, solid, dot-dashed, dashed)
with $\mu=0$, $\sigma=1$ and $a=0.01$.
\label{curtose}}
\end{center}
\end{figure}

The skewness and kurtosis measures can now be calculated from the ordinary moments
using well-known relationships.
Plots of the skewness and kurtosis
are shown in Figures~\ref{assimetria} and~\ref{curtose}, respectively.
The curves are given
for some choices of $b$ as functions of $a$ and $c$
and
for some choices of $a$ as functions of $b$ and $c$,
respectively.
These
figures immediately reveal that
the skewness and kurtosis are
very flexible for different values of $a$, $b$, and $c$.

\section{Generating Function}
\label{section.generating}

In this section, we provide a representation for the mgf of the
$\operatorname{McN}(a,b,0,1)$ distribution, say $M(-s)=\exp(-s Z)$.
From equation (\ref{exppdf2}), we obtain
\begin{eqnarray*}
M(-s)=\frac{1}{\sqrt{2\pi}}\sum_{k,r=0}^\infty t_k\,s_r((k+a)c-1)\,\int_{-\infty}^{\infty}\Phi(x)^r\,
\exp\left(-s x-\frac{x^2}{2}\right)\mathrm{d}x.
\end{eqnarray*}

The standard normal cdf $\Phi(x)$ can be expressed
as a power series expansion
$\Phi(x)=\sum_{j=0}^{\infty}a_j \, x^{j}$,
where
$a_0=(1+\sqrt{2/\pi})^{-1}/2$,
$a_{2j+1}=\frac{(-1)^{j}2^{-j}}{\sqrt{2\pi}(2j+1)j!}$
for $j=0,1,2\ldots$,
and
$a_{2j}=0$ for $j=1,2,\ldots$
Thus, we can write
\begin{align}
\label{powercdf}
\Phi(x)^r=\sum_{j=0}^{\infty}c_{r,j}\, x^{j},
\end{align}
where the coefficients $c_{r,j}$ are calculated from the recurrence
equation (\ref{formulae_c}) with the $a_i'$s given before.
We have
\begin{eqnarray*}
M(-s)=\frac{1}{\sqrt{2\pi}}\sum_{k,r,j=0}^\infty t_k\,s_r((k+a)c-1)\,c_{r,j}
\int_{-\infty}^{\infty}\,x^{j}\,
\exp\left(-s x-\frac{x^2}{2}\right)\mathrm{d}x.
\end{eqnarray*}
By making use of a result
by~\shortciteauthor{prudnikov1986integrals}~\citeyear[Eq.~2.3.15.8]{prudnikov1986integrals},
the integral follows as
\begin{eqnarray}\label{basicintegral}
J(s,j)
=
\int_{-\infty}^{\infty}\,x^{j}\,
\exp\left(-s x-\frac{x^2}{2}\right)\mathrm{d}x
=
(-1)^j
\,
\sqrt{2\pi}
\,
\frac{\partial^j}{\partial s^j}
\left\{\exp\left(\frac{s^2}{2}\right)\right\}
\end{eqnarray}
and thus
\begin{eqnarray}\label{eqmgf}
M(-s)=\frac{1}{\sqrt{2\pi}}
\sum_{k,r,j=0}^\infty t_k\,s_r((k+a)c-1)\,c_{r,j}\,
J(s,j).
\end{eqnarray}
Equation (\ref{eqmgf}) is the main result of this section. The
characteristic function (chf) $\phi(t)=\operatorname{E}[\exp(\imath tX)]$ of the standard
McN distribution is immediately obtained by $\phi(-s)=M(- \imath s)$, where $\imath=\sqrt{-1}$.

\section{Mean Deviations}
\label{section.meandeviations}

Let $Z\sim \operatorname{McN}(a,b,0,1,)$.
The amount of scatter in $Z$ is measured
to some extent by the totality of deviations from the mean and median.
These are known as the mean deviations about
the mean and median,
defined by
\begin{eqnarray*}
\delta_1 (Z) = \int_0^{\infty }\left | x - \mu_1^{\prime} \right | f(x) \mathrm{d}x
\quad
\text{and}
\quad
\delta_2 (Z) = \int_0^{\infty} \left | x - m \right | f(x) \mathrm{d}x,
\end{eqnarray*}
respectively, where $\mu_1^{\prime}=\operatorname{E}(Z)$ and
$m$ is the median of $Z$.
The measures $\delta_1 (Z)$ and $\delta_2 (Z)$ can be expressed as
\begin{eqnarray}\label{deltas}
\delta_1(Z)=2\mu_1^{\prime}F(\mu_1^{\prime})- 2 P(\mu_1^{\prime})
\quad
\text{and}
\quad
\delta_2(Z)=\mu_1^{\prime}- 2 P(m),
\end{eqnarray}
where $P(q)=\int_0^q x f(x)\mathrm{d}x$.
Combining (\ref{exppdf2}) and (\ref{powercdf}), we obtain
\begin{eqnarray*}
P(q)=\frac{1}{\sqrt 2\pi}\sum_{k,r,j=0}^\infty t_k\,c_{r,j}\,s_r((k+a)c-1)\,\int_{-\infty}^{q} x^{j+1}\,
\exp\left(-\frac{x^2}{2}\right)\mathrm{d}x,
\end{eqnarray*}
where the coefficients $c_{r,j}$ are obtained from (\ref{formulae_c})
from the $a_i'$s just given before (\ref{powercdf}).

We can determine $A(j,q)=\int_{-\infty}^{q} x^{j}\,\exp\left(-\frac{x^2}{2}\right)\mathrm{d}x$
for $q<0$ and $q>0$. Let
$$G(j)=\int_{0}^{\infty} x^{j}\exp\left(-\frac{x^2}{2}\right)\mathrm{d}x=2^{(j-1)/2}\,\Gamma\left(\frac{j+1}{2}\right).$$
For $q<0$,
$$A(j,q)=(-1)^{j}G(j)+ (-1)^{j+1}H(j,q),$$
whereas, for $q>0$,
$$A(j,q)= (-1)^{j}G(j)+ H(j,q),$$
where the integral $H(j,q)=\int_{0}^{j}x^j \exp\left(-\frac{x^2}{2}\right)\mathrm{d}x$
can be easily determined as \cite{Whittaker1990}
\begin{eqnarray*}
H(j,q)&=&\frac{2^{j/4+1/4} q^{j/2+1/2} e^{-q^2/4}}{(j/2+1/2)(j+3)}M_{j/4+1/4,j/4+3/4}(q^{2}/2)\\ \nonumber
&&+\frac{2^{j/4+1/4} q^{j/2-3/2} e^{-q^{2}/4}}{j/2+1/2}M_{j/4+5/4,j/4+3/4}(q^2/2),
\end{eqnarray*}
where
$M_{k,m}(x)$ is the Whittaker function.
This function can be expressed in
terms of the confluent hypergeometric function
$${_1F}_{1}(a;b;z)=\sum\limits_{k=0}^{\infty}\frac{(a)_{k}}{(b)_{k}}\frac{z^{k}}{k!}$$
where $(a)_{k}=a\cdot (a-1) \cdots (a-k+1)$
is the ascending factorial
(with the convention that $(a)_{0}=1$),
by
\begin{align*}
M_{k,m}(x)
=
\exp\left(-\frac{x}{2}\right)
x^{m+1/2}
{_1F}_{1}
\left(
\frac{1}{2}+m-k;1+2m;x
\right)
.
\end{align*}

Hence, we have all quantities to obtain
\begin{eqnarray}\label{basicmean}
P(q)=\frac{1}{\sqrt 2\pi}\sum_{k,r,j=0}^\infty t_k\,c_{r,j}\,s_r((k+a)c-1)\,A(j+1,q),
\end{eqnarray}
where $A(j+1,q)$ was defined before. Equations~\eqref{deltas} and~\eqref{basicmean}
give the mean deviations.

\section{Order statistics}
\label{section.orderstatistics}

The density function $f_{i:n}(x)$ of the $i$th order statistic for
$i=1,\ldots,n$ from data values $X_1,\ldots,X_n$ following the standard
McN distribution can be expressed as
\begin{eqnarray}\label{pdforder}
f_{i:n}(x)=
\frac{f(x)}{\operatorname{B}(i,n-i+1)}\sum_{j=0}^{n-i}(-1)^j\binom{n-i}{j}F(x)^{i+j-1}.
\end{eqnarray}

We can use the incomplete beta function expansion for $\beta$ real non-integer
\begin{eqnarray*}
I_x (a,b)=\frac{x^{a}}{\operatorname{B}(a,b)}\sum_{m =0}^\infty \frac{(1-b)_m x^m}{(a+ m)m!},
\end{eqnarray*}
where $(f)_k=\Gamma(f+k)/\Gamma(f)$.
It follows from equations~\eqref{cdf} and~\eqref{powerG} that
\begin{eqnarray*}
F(x)=\frac{1}{\operatorname{B}(a,b)}\sum_{m,p=0}^\infty \frac{(1-b)_m\, s_p((a+m)c)}{(a+ m)\,m!} \, \Phi(x)^p,
\end{eqnarray*}
and then
\begin{eqnarray*}
F(x)=\sum_{r=0}^\infty v_p\,\Phi(x)^p,
\end{eqnarray*}
where
$$v_p=v_p(a,b)=\frac{1}{\operatorname{B}(a,b)}\sum_{m=0}^\infty \frac{(1-b)_m\, s_p((a+m)c)}{(a+ m)m!}.$$
Thus, the $(i+j-1)$th power of $F(x)$ can be determined as
\begin{eqnarray*}
F(x)^{i+j-1}=\left(\sum_{p=0}^\infty v_p\,\Phi(x)^p\right)^{i+j-1}=\sum_{p=0}^\infty d_{i+j-1,p}\,\Phi(x)^p,
\end{eqnarray*}
where the coefficients $d_{n,p}$
are
obtained from the recurrence equation
\begin{align*}
d_{n,p}
=
\frac{1}{r v_0}
\sum_{m=1}^{p}
[m(n+1)-p]\,v_m\,d_{n,p-m},
\quad
p=1,2,\ldots,
\end{align*}
where $d_{n,0}=v_0^n$. The density function~\eqref{pdforder} can be
rewritten as
\begin{eqnarray}\label{pdforder1}
f_{i:n}(x)
=
\phi(x)
\sum_{p,r=0}^\infty
g_{i:n}(p,r)\,\Phi(x)^{p+r}
,
\end{eqnarray}
where the coefficient $g_{i:n}(p,r)$ is given by
\begin{align*}
g_{i:n}(p,r)
=
g_{i:n}(p,r)(a,b,c)
=
\frac{1}{\operatorname{B}(i,n-i+1)}
\sum_{k=0}^\infty
t_k\, s_r((k+a)c-1)
\sum_{j=0}^{n-i}(-1)^j
\binom{n-i}{j}d_{i+j-1,p}
.
\end{align*}
Equation~\eqref{pdforder1} is the main result of this section.
It gives the
density function of the McN order statistics
as a power series of the standard normal cumulative function multiplied by
the standard normal density function.

\section{Properties of order statistics}
\label{section.propertiesorderstatistics}

Here, we provide two expansions for the moments
and one expansion for the mgf of the McN order statistics.
First, the $n$th moment of the $i$th order
statistic in a sample of size $n$, say $\operatorname{E}(X_{i:n}^{n})$, of the
McN distribution follows from equation (\ref{pdforder1})
\begin{eqnarray}\label{morder1}
\operatorname{E}(X_{i:n}^{n})=\sum_{p,r=0}^\infty g_{i:n}(p,r)\,\tau_{n,p+r},
\end{eqnarray}
where $\tau_{n,p+r}$ is easily obtained from (\ref{pwmnormal}).
Then, the ordinary moments of the McN order statistics
are simple linear functions of the PWMs of
the normal distribution. An alternative
formula can be immediately derived from (\ref{pdforder1}) by comparing
equations (\ref{moment1}) and (\ref{moment2}). We have
\begin{eqnarray}\label{morder2}
\operatorname{E}(X_{i:n}^{n})=\sum_{p,r,m=0}^\infty g_{i:n}(p,r)\, (2\pi)^{m/2}\,c_{n,m}\,\sum_{j=0}^{m}\frac{(-1)^j 2^{-j}\binom{m}{j}}{p+r+1+m-j}.
\end{eqnarray}

The mgf of the $i$th order statistic from
the McN distribution, say $M_{i:n}(s)$,
can be written from~\eqref{pdforder1} as
\begin{eqnarray*}
M_{i:n}(-s)=\frac{1}{\sqrt 2\pi}\sum_{p,r=0}^\infty g_{i:n}(p,r)\,\int_{-\infty}^{\infty}\Phi(x)^{p+r}
\exp\left(-s x-\frac{x^2}{2}\right)\mathrm{d}x
.
\end{eqnarray*}
Using the same algebraic
development of Section~\ref{section.generating},
we obtain
\begin{eqnarray}
\label{mgforderstatistic}
M_{i:n}(-s)
=
\frac{1}{\sqrt 2\pi}
\sum_{p,r,j=0}^\infty
g_{i:n}(p,r)\,c_{p+r,j}\,J(s,j),
\end{eqnarray}
where $J(s,j)$ is defined by~\eqref{basicintegral}.
Equations~\eqref{morder1}, \eqref{morder2} and \eqref{mgforderstatistic}
are the main results of this section.

\section{Hazard Function}

\label{section.hazard}

%As an important tool to characterize lifetime phenomena, t
The McN hazard rate function
takes the form
\begin{align*}
h(x)
%&\triangleq
%\frac{f(x)}{1-F(x)}
%\\
%&
=
\frac{c\left[\Phi\left(\frac{x-\mu}{\sigma}\right)
\right]^{ac-1}\,
\left[1-\Phi^c\left(\frac{x-\mu}{\sigma}\right)
\right]^{b-1}\,\phi\left(\frac{x-\mu}{\sigma}\right)}
{\sigma\,\operatorname{B}(a,b)
\left[1-I_{\Phi^c(\frac{x-\mu}{\sigma})}(a,b)
\right]}
.
\end{align*}
We study the asymptotic behavior of the hazard function as $x\to\infty$.
%Let $h_{\text{N}}(x)$ be the hazard rate function of the normal distribution. We show that $h_{\text{N}}(x)\sim x/\sigma^2$ as $x\to\infty$. We have:
%\begin{align*}
%\lim_{x\to\infty}
%\frac{h_{\text{N}}(x)}{x}
%&=
%\lim_{x\to\infty}
%\frac{\phi(\frac{x-\mu}{\sigma})\frac{1}{x}}{\sigma[1-\Phi(\frac{x-\mu}{\sigma})]}.
%\end{align*}
%Direct evaluation of this limit gives an indeterminate form. However, using L'H\^opital rule, straightforward manipulations show that the above limit is simply $1/\sigma^2$.
We now show that
$h(x)\sim (b/\sigma^2) x$
as $x\to\infty$.
Indeed, we have:
\begin{align*}
\lim_{x\to\infty}
\frac{h(x)}{x}
&=
\frac{c}{\operatorname{B}(a,b)\sigma}
\lim_{x\to\infty}
\left[
\Phi\left(\frac{x-\mu}{\sigma}\right)
\right]^{ac-1}
\lim_{x\to\infty}
\frac{
\left[
1-\Phi^c\left(\frac{x-\mu}{\sigma}\right)
\right]^{b-1}
\phi\left(\frac{x-\mu}{\sigma}\right)\frac{1}{x}}
{1-I_{\Phi^c(\frac{x-\mu}{\sigma})}(a,b)}
\\
&=
\frac{c}{\operatorname{B}(a,b)\sigma}
\lim_{x\to\infty}
\frac{\left[
1-\Phi^c\left(\frac{x-\mu}{\sigma}\right)
\right]^{b-1}
\phi\left(\frac{x-\mu}{\sigma}\right)
\frac{1}{x}}{1-I_{\Phi^c(\frac{x-\mu}{\sigma})}(a,b)}.
\end{align*}
Once again, another indeterminate form arises. An application of L'H\^opital's rule yields
%\begin{align*}
%&\lim_{x\to\infty}
%\frac{
%\left[
%1-\Phi\left(\frac{x-\mu}{\sigma}\right)
%\right]^{b-1}
%\phi\left(\frac{x-\mu}{\sigma}\right)\frac{1}{x}}
%{1-I_{\Phi(\frac{x-\mu}{\sigma})}(a,b)}
%\\
%&=
%\lim_{x\to\infty}
%\frac{(b-1)
%\left[
%1-\Phi\left(\frac{x-\mu}{\sigma}\right)
%\right]^{b-2}
%(-1)[\phi\left(\frac{x-\mu}{\sigma}\right)]^2\frac{1}{\sigma}\frac{1}{x}}
%{(-1)(\operatorname{B}(a,b))^{-1}\left[\Phi\left(\frac{x-\mu}{\sigma}\right)
%\right]^{a-1}\left[1-\Phi\left(\frac{x-\mu}{\sigma}\right)
%\right]^{b-1}\phi\left(\frac{x-\mu}{\sigma}\right)\frac{1}{\sigma}}
%\\
%&+\lim_{x\to\infty}\frac{[1-\Phi\left(\frac{x-\mu}{\sigma}\right)
%]^{b-1}\phi\left(\frac{x-\mu}{\sigma}\right)(-1)\frac{1}{\sigma}(\frac{x-\mu}{\sigma})\frac{1}{x}}{(-1)(\operatorname{B}(a,b))^{-1}\left[\Phi\left(\frac{x-\mu}{\sigma}\right)
%\right]^{a-1}\left[1-\Phi\left(\frac{x-\mu}{\sigma}\right)
%\right]^{b-1}\phi\left(\frac{x-\mu}{\sigma}\right)\frac{1}{\sigma}
%}
%\\
%&+\lim_{x\to\infty}\frac{[1-\Phi\left(\frac{x-\mu}{\sigma}\right)
%]^{b-1}\phi\left(\frac{x-\mu}{\sigma}\right)(-1)\frac{1}{x^2}}
%{
%(-1)(\operatorname{B}(a,b))^{-1}\left[\Phi\left(\frac{x-\mu}{\sigma}\right)
%\right]^{a-1}\left[1-\Phi\left(\frac{x-\mu}{\sigma}\right)
%\right]^{b-1}\phi\left(\frac{x-\mu}{\sigma}\right)\frac{1}{\sigma}
%}
%\end{align*}
%
%Since the last limit is equal to zero, we obtain
\begin{align*}
\lim_{x\to\infty}
&
\frac{
\left[
1-\Phi^c\left(\frac{x-\mu}{\sigma}\right)
\right]^{b-1}
\phi\left(\frac{x-\mu}{\sigma}\right)\frac{1}{x}}
{1-I_{\Phi^c(\frac{x-\mu}{\sigma})}(a,b)}
\\
%&=
%\sigma\operatorname{B}(a,b)
%\left\{
%\lim_{x\to\infty}
%\frac{
%(b-1)
%\left[
%1-\Phi\left(\frac{x-\mu}{\sigma}\right)
%\right]^{-1}
%\phi\left(\frac{x-\mu}{\sigma}\right)\frac{1}{\sigma}\frac{1}{x}}{\left[\Phi\left(\frac{x-\mu}{\sigma}\right)
%\right]^{a-1}}
%+\lim_{x\to\infty}\frac{\frac{1}{\sigma}(\frac{x-\mu}{\sigma})\frac{1}{x}}
%{
%\left[\Phi\left(\frac{x-\mu}{\sigma}\right)
%\right]^{a-1}
%}
%\right\}
%\\
&
=
%\sigma\operatorname{B}(a,b)
%\left\{
%\lim_{x\to\infty}
%\frac{(b-1)h_{\text{N}}(x)\frac{1}{x}}{\left[\Phi\left(\frac{x-\mu}{\sigma}\right)
%\right]^{a-1}}+\lim_{x\to\infty}\frac{\frac{1}{\sigma}(\frac{x-\mu}{\sigma})\frac{1}{x}}{\left[\Phi\left(\frac{x-\mu}{\sigma}\right)
%\right]^{a-1}}
%\right\}
%\\
%&
%=
\frac{\sigma\operatorname{B}(a,b)}{c}
\left\{
\frac{b-1}{\sigma}
\lim_{x\to\infty}
\frac{\phi(\frac{x-\mu}{\sigma})\frac{1}{x}}
{1-\Phi^c(\frac{x-\mu}{c})}
+
\frac{1}{c\sigma^2}
\lim_{x\to\infty}
\frac{x-\mu}{x}
\right\}.
%=
%\operatorname{B}(a,b)\frac{b}{\sigma}.
\end{align*}

Another application of L'H\^opital's rule gives
\begin{align*}
\frac{\sigma\operatorname{B}(a,b)}{c}
\left\{
\frac{b-1}{\sigma}
\lim_{x\to\infty}
\frac{\phi(\frac{x-\mu}{\sigma})\frac{1}{x}}
{1-\Phi^c(\frac{x-\mu}{c})}
+
\frac{1}{\sigma^2}
\lim_{x\to\infty}
\frac{x-\mu}{x}
\right\}
=
%\frac{\sigma\operatorname{B}(a,b)}{c}
%\frac{b}{\sigma^2}
\frac{b\operatorname{B}(a,b)}{c\sigma}
\lim_{x\to\infty}\frac{x-\mu}{x}
=
%\frac{b}{c\,\sigma}
%\operatorname{B}(a,b)
\frac{b\operatorname{B}(a,b)}{c\sigma}
.
\end{align*}

Thus,
$h(x)\sim (b/\sigma^2)x$ as $x\to\infty$.
It is worth noting that this asymptotic behavior does not depend on the parameters $a,c$ and $\mu$.
The limit of $h(x)$ as $x\to-\infty$ is zero.
However, we can verify that
$h(x)\sim
\frac{c}{\sigma\operatorname{B}(a,b)}
\left(-\frac{x-\mu}{\sigma}\right)^{1-ac}
\left[\phi(\frac{x-\mu}{\sigma})\right]^{ac}$
when $x\to-\infty$.
In fact, considering that
$\lim_{x\to-\infty} \Phi(\frac{x-\mu}{\sigma})=\lim_{x\to-\infty} I_{\Phi(\frac{x-\mu}{\sigma})}(a,b)=0$,
we obtain
\begin{align*}
\lim_{x\to-\infty}
%&
\frac{h(x)}
{
\left(-\frac{x-\mu}{\sigma}\right)^{1-ac}
\left[\phi(\frac{x-\mu}{\sigma})\right]^{ac}
}
=
%\\
%&
\frac{c}{\sigma\operatorname{B}(a,b)}
\lim_{x\to-\infty}
\left[
\Phi\left(\frac{x-\mu}{\sigma}\right)
\right]^{ac-1}
\frac{\phi\left(\frac{x-\mu}{\sigma}\right)}{
  \left(
    -\frac{x-\mu}{\sigma}
  \right)^{1-ac}
  \left[
    \phi(\frac{x-\mu}{\sigma})
  \right]^{ac}
}
.
\end{align*}
Applying L'H\^{o}pital rule,
we can show that $\Phi(x)\sim (-x)^{-1}\phi(x)$.
So, we obtain
that the above limit is $c/[\sigma\operatorname{B}(a,b)]$.

%Leandro1: Falta adicionar esta referência.
\shortciteauthor{rego2012properties}~\citeyear{rego2012properties}
demonstrated
that $h_{\text{N}}(x)
\sim
\phi\left(\frac{x-\mu}{\sigma}\right)
/
\sigma$ as $x\to-\infty$,
where $h_{\text{N}}(x)$ is the normal hazard rate function.
Hence,
the asymptotic behavior of $h(x)$ can be related to
that one of $h_{\text{N}}(x)$ when $x\to-\infty$ by

\begin{align*}
h(x)
\sim
\frac{c}{\operatorname{B}(a,b)\sigma^{1-ac}}
\left(
-\frac{x-\mu}{\sigma}
\right)^{1-ac}
\left[
h_{\text{N}}(x)
\right]^{ac}
.
\end{align*}

\section{Shannon Entropy}

\label{section.entropy}

Shannon entropy is a measure of uncertainty associated with a random variable.
Consider a random variable $X\sim \mathrm{McN}(a,b,c,\mu,\sigma^2)$. Thus,
the Shannon entropy of $X$ is given by
\begin{align*}
\mathrm{H}(X)
=&
-
\operatorname{E}\{\log \left[f(X) \right]\}
=
-
\int_{-\infty}^{\infty} f(x)\log \left[f(X) \right] \mathrm{d}x
\\
%=
%&
%-\int_{\infty}^{\infty}
%f(x)
%\log
%\left\{
%\frac{1}{\sigma\mathrm{B}(a,b)}
%\left[
%\Phi\left(\frac{x-\mu}{\sigma} \right)
%\right]^{a-1}
%\left[
%1 - \Phi\left( \frac{x-\mu}{\sigma} \right)
%\right]^{b-1}
%\phi\left( \frac{x-\mu}{\sigma} \right)
%\right\}
%\mathrm{d}x
%\\
=
&
-
\int_{-\infty}^{\infty}
f(x)
\log
\left\{
\frac{c}{\sigma\mathrm{B}(a,b)}
\right\}
\mathrm{d}x
-
(ac-1)
\int_{-\infty}^{\infty}
f(x)
\log
\left\{
\Phi\left( \frac{x-\mu}{\sigma}\right)
\right\}
\mathrm{d}x
\\
&
-
(b-1)
\int_{-\infty}^{\infty}
f(x) 
\log
\left\{
1-\Phi^c\left( \frac{x-\mu}{\sigma}\right)
\right\}
\mathrm{d}x
-
\int_{-\infty}^{\infty}f(x)
\log
\left\{
\phi\left( \frac{x-\mu}{\sigma} \right)
\right\}
\mathrm{d}x
.
\end{align*}
The first integral is simply equal to
$-\log[\sigma c^{-1}\mathrm{B}(a,b)]$.
The second and third integrals can be expressed in terms of the beta function.
Indeed,
using Taylor series expansions for the logarithmic function,
%we obtain
%\begin{align*}
%&\log\left\{
%\Phi\left( \frac{x-\mu}{\sigma}\right)\right\}=
%\sum_{n=1}^{\infty}\frac{(-1)^n}{n}\,\left[
%1-\Phi\left( \frac{x-\mu}{\sigma}\right)\right]^n
%\end{align*}
%and
%\begin{align*}
%\log\left\{1-\Phi\left( \frac{x-\mu}{\sigma}\right)\right\}
%=\sum_{n=1}^{\infty}\frac{(-1)^n}{n}\,
%\Phi\left( \frac{x-\mu}{\sigma}\right)^n.
%\end{align*}
%As a consequence,
the previous integrals reduce to
$$
(ac-1)\int_{-\infty}^{\infty}
f(x)
\log\left\{
\Phi\left(\frac{x-\mu}{\sigma}\right)
\right\}
\mathrm{d}x
=
\frac{1-ac}{\mathrm{B}(a,b)}
\sum_{n=1}^{\infty}
\sum_{k=0}^{n}\frac{(-1)^{k}\,\binom{n}{k}}{n}
\mathrm{B}\left(\frac{ac+k}{c},b\right)
$$
and
$$
(b-1)\int_{-\infty}^{\infty}
f(x)
\log\left\{1-\Phi^c\left(\frac{x-\mu}{\sigma}\right)
\right\}
\mathrm{d}x
=
\frac{1-b}{\mathrm{B}(a,b)}
\sum_{n=1}^{\infty}\frac{\mathrm{B}(n+a,b)}{n}
.
$$
Now, we can write the last integral as
\begin{align*}
\int_{-\infty}^{\infty}f(x)\,\log\left\{
\phi\left( \frac{x-\mu}{\sigma} \right)
\right\}
\mathrm{d}x
&
=
\int_{-\infty}^{\infty}
f(x)
\left[
\log
\left(
\frac{1}{\sqrt{2\pi}}
\right)
+
\log\left\{
\exp\left[-\frac{(x-\mu)^2}{2\sigma^2}\right]
\right\}
\right]
\mathrm{d}x
\\
&
=
-\log\left(\sqrt{2\pi}\right)
-
\frac{1}{2\sigma^2}
\left[
\operatorname{E}(X^2)-2\mu \operatorname{E}(X) + \mu^2
\right]
,
\end{align*}
where $\operatorname{E}(X^2)$ and $\operatorname{E}(X)$ follow from
(\ref{moment2})
and (\ref{moment3}).
Finally, we obtain
\begin{align*}
\mathrm{H}(X)
=&
\log\left\{
\sqrt{2\pi} \sigma c^{-1}\mathrm{B}(a,b)
\right\}
+
\frac{1}{2\sigma^2}
\left[
\operatorname{E}(X^2)-2\mu \operatorname{E}(X) + \mu^2
\right]
\\
&
+
%K\sigma
\frac{1}{\mathrm{B}(a,b)}
\sum_{n=1}^{\infty}
\frac{1}{n}
\left[(b-1)\mathrm{B}(n+a,b)
+
(ac-1)
\sum_{k=0}^{n}(-1)^{k}
\binom{n}{k}
\mathrm{B}\left(\frac{ac+k}{c},b\right)
\right]
.
\end{align*}

\begin{figure}
\centering
\subfigure[]{\epsfig{file=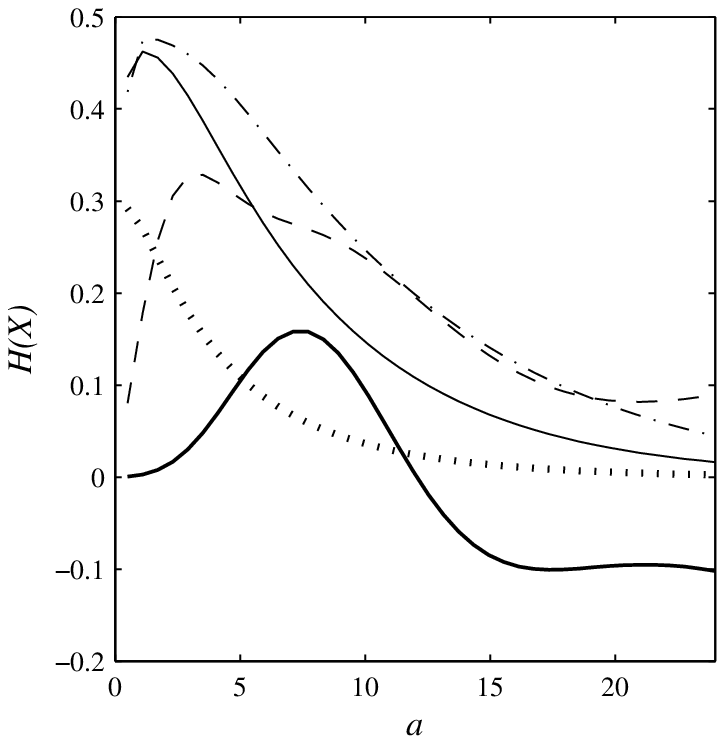}}
\quad
\subfigure[]{\epsfig{file=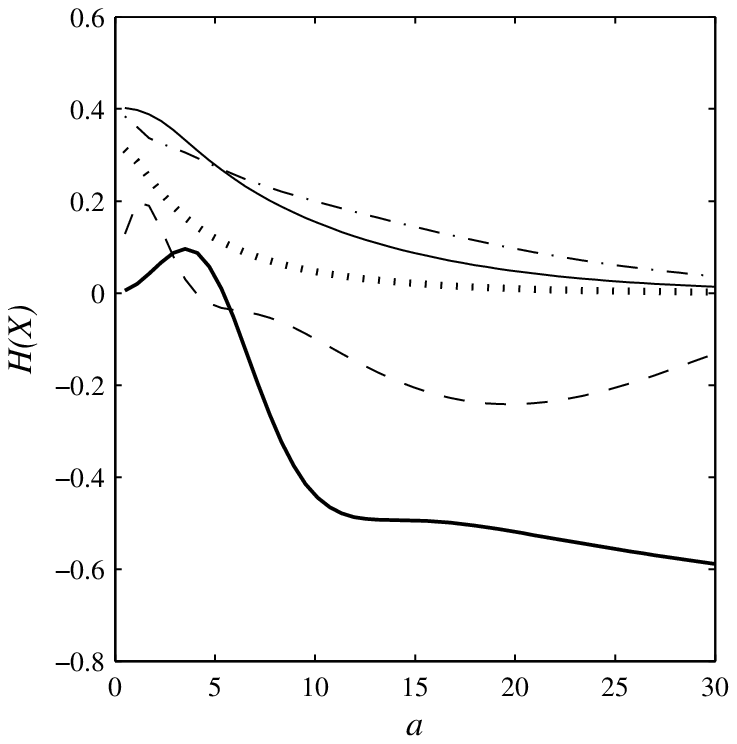}}
\caption{(a) $\mu=0$, $\sigma=1$, $b\in\{ 0.1, 0.5, 1, 5, 15, 20\}$, $c=0.75$;
(b) $\mu=0$, $\sigma=1$, $b\in\{ 0.1, 0.5, 1, 5, 15, 20\}$, $c=5$.}
\label{fig7}
\end{figure}

Figure~\ref{fig7} depicts some plots of the Shannon entropy for selected pdf parameters.

\section{Estimation} \label{sec:est}

Here,
we determine the maximum likelihood estimates (MLEs)
of the parameters of the
$\operatorname{McN}(a,b,c,\mu,\sigma^2)$
distribution from complete samples only.
Let $x_{1},\ldots,x_{n}$ be a random sample of size $n$ from this
distribution.
The log-likelihood function for the vector of
parameters $\bm{\theta}=(a,b,c,\mu,\sigma)^{T}$
reduces to
\begin{align*}
l(\bm{\theta})
=
&
n
\log(c)
-
n
\log
\left[\operatorname{B}(a,b)\right]
-
+\sum_{i=1}^n
\log
\left[
\phi(z_i)
\right]
+
(ac-1)
\sum_{i=1}^n
\log
\left[
\Phi(z_i)
\right]
\\
&
+
(b-1)
\sum_{i=1}^n
\log\left[1-\Phi^{c}(z_i)\right]
,
\end{align*}
where $z_{i}=\frac{x_{i}-\mu}{\sigma}$.
The components of the score vector $U(\bm{\theta})$ are given by
\begin{align*}
U_{a}(\bm{\theta})
=
&
-n
\left[
\psi^{(0)}(a)+\psi^{(0)}(a+b)
\right]
+
c
\sum_{i=1}^n
\log
\left[\Phi(z_i)
\right]
,
\\
U_{b}(\bm{\theta})
=
&
-n
\left[
\psi^{(0)}(a)+\psi^{(0)}(a+b)
\right]
+
\sum_{i=1}^n
\log
\left[1 - \Phi^c(z_i)
\right]
,
\\
U_{c}(\bm{\theta})
=
&
\frac{n}{c}
+
a
\sum_{i=1}^n
\log
\left[\Phi(z_i)
\right]
-
(b-1)
\sum_{i=1}^n
\frac{\Phi^{c}(z_i)\log\left[\Phi(z_i)\right]
}{1-\Phi^{c}(z_i)},
\\
U_{\mu}(\bm{\theta})
=
&
\frac{1}{\sigma}
\sum_{i=1}^n
z_i
-
\frac{ac-1}{\sigma}
\sum_{i=1}^n
\frac{\phi(z_i)}{\Phi(z_i)}
+
\frac{(b-1)c}{\sigma}
\sum_{i=1}^n
\frac{\phi(z_i)
\Phi^{c-1}(z_i)}{1-\Phi^{c}(z_i)}
,
\\
U_{\sigma}(\bm{\theta})
=&
\frac{1}{\sigma}
\sum_{i=1}^n
z_i^2
-
\frac{ac-1}{\sigma}
\sum_{i=1}^n
\frac{z_i\phi(z_i)}{\Phi(z_i)}
+
\frac{(b-1)c}{\sigma}
\sum_{i=1}^n
\frac{z_i\phi(z_i)
\Phi^{c-1}(z_i)}{1-\Phi^{c}(z_i)}
,
\end{align*}
where $\psi^{(0)}(\cdot)$ is the digamma function.
Setting these expressions to zero and solving them simultaneously
yields the MLEs of the five parameters.
For interval estimation on the model parameters, we require the expected
information matrix. The elements of the $5\times 5$ unit observed information
matrix $\mathcal{I}(\bm{\theta})$ are given in Appendix~\ref{section.appendix}.
Under conditions that are
fulfilled for parameters in the interior of the parameter space but not
on the boundary, the asymptotic distribution of
\begin{align*}
\sqrt{n}
(\widehat{\bm{\theta}}-\bm{\theta})
\quad
\text{is}
\quad
N_5(0,K(\bm{\theta})^{-1})
,
\end{align*}
where $K(\bm{\theta})=\operatorname{E}\{\mathcal{I}(\bm{\theta})\}$
is the expected information matrix.

The multivariate normal $N_5(0, \mathcal{I}(\bm{\theta})^{-1})$ distribution can be
used to construct approximate confidence intervals and confidence regions
for the individual parameters.

The likelihood ratio (LR) can be used for testing goodness of fit of
the McN distribution and for comparing this distribution with some
of its sub-models. We can compute the maximum values of the
unrestricted and restricted log-likelihoods to construct LR statistics
for testing some sub-models of the McN distribution. For example,
we may use the LR statistic to check whether the fit using the McN
distribution is statistically ``superior'' to a fit using the
exponentiated normal (EN) and normal distributions for a given data set.
In any case, hypothesis tests of the type $\mathcal{H}_{0}:\bm{\psi}=\bm{\psi}_{0}$ versus $\mathcal{H}_1:\bm{\psi} \neq \bm{\psi}_{0}$,
where $\bm{\psi}$ is a vector formed with some components of $\bm{\theta}$
and $\bm{\psi}_{0}$ is a specified vector, can be performed using
LR statistics. For example, the test of $\mathcal{H}_{0}:a=b=1$ versus $\mathcal{H}_1:\mathcal{H}_{0}
\text{\space\emph{is not true}}$ is equivalent to compare the McN distribution
with the normal distribution and the LR statistic reduces to
\begin{align*}
w
=
2
\{
\ell(\widehat{a},\widehat{b},\widehat{c},\widehat{\mu},\widehat{\sigma})
-
\ell(1,1,1,\widetilde{\mu},\widetilde{\sigma})
\}
,
\end{align*}
where $\widehat{a}$, $\widehat{b}$, $\widehat{c}$, $\widehat{\mu}$, and
$\widehat{\sigma}$ are the MLEs under $\mathcal{H}$
and $\widetilde{\mu}$ and $\widetilde{\sigma}$ are the estimates under
$\mathcal{H}_{0}$.

\section{Applications}
\label{section.applications}

In this section,
 we use three real data sets to compare the fits of a McN distribution with some of its sub-models, i.e., the BN, KwN, EN, normal and skew-normal distributions.
In each case,
the parameters are estimated by maximum likelihood
(Section~\ref{sec:est})
using the SAS subroutine NLMixed.
Convergence was achieved using the re-parametrization $a^\ast=ac$.

First, we describe the data sets.
Then, we report the MLEs (and the corresponding standard errors in parentheses) of the parameters and the values of the AIC (Akaike Information Criterion), CAIC (Consistent Akaike Information Criterion) and BIC (Bayesian Information Criterion) statistics. The lower the values of these criteria, the better the fit. Next, we perform LR tests for the need of skewness parameters. Finally, histograms of the data sets are provided
to obtain a visual comparison of the fitted density functions.

Consider the data set discussed in
in \citeauthor{Weisberg2005}~\citeyear[Section 6.4]{Weisberg2005}
which represents 102 male and 100 female athletes collected at the Australian Institute of Sport, courtesy of Richard Telford and Ross Cunningham. The following variables are evaluated in this study:
\begin{enumerate}
  \item Plasma ferritin concentration (plasma);
  \item Sum of skin folds (skin folds);
  \item Lean body mass (mass).
\end{enumerate}

\begin{table}
\centering {\caption{Descriptive statistics\label{summary}}
\vspace*{0.3cm}
\begin{tabular}{ccccccccccc}
\hline
Data  & Mean & Median & Mode & SD  & Variance & Skewness & Kurtosis& Min.& Max.\\
\hline
Plasma  & 76.88  &65.50  & 41 & 47.50 & 2256.37 & 1.29 &1.49  & 8 & 234 \\
\hline
Skin folds  &  69.02 & 58.6 & 52.8$^q$ & 32.57 & 1060.50 & 1.18 & 1.43 & 28 &200.8\\
\hline
Mass  & 64.87  & 63.04 & 78 & 13.07 & 170.83 & 0.36 & -0.22 & 34.36 &106\\
   \hline
\end{tabular}
\\$^{q}$ there are several modes.
}
\end{table}

Table \ref{summary} gives a descriptive summary of each sample. The plasma and skin folds have positive skewness and kurtosis, larger values of these sample moments being apparent in the plasma data. The mass data has positive skewness and negative kurtosis.

We now compute the MLEs and the AIC, BIC and CAIC statistics for each data set.
For the three data sets,
we fit the McN distribution, with parameters $a$, $b$, $c$, $\mu$ and $\sigma$, and this is compared with the fits obtained using the BN, KwN, EN, normal and skew normal distributions. The estimates of $\mu$ and $\sigma$ for the normal distribution were adopted as initial values.

\begin{table}[!htb]
\centering {\caption{MLEs and information criteria.
\label{EMV}}
\vspace*{0.3cm}
\begin{tabular}{c|ccccc|ccc}
\hline
\hline \textbf{Plasma}  & $a$ & $b$ &$c$ & $\mu$ & $\sigma$  &$\rm{AIC}$ &$\rm{BIC}$&$\rm{CAIC}$\\
\hline
McN  &  261.74  & 0.2112 & 0.0032 & -93.9830& 43.7127 &  2071.4 & 2071.8 & 2088.0 \\
 & (5.6963)& (0.0263)& (0.00007)& (8.2467) & (0.8559)& &&   \\\hline
BN  &  8.6410  & 0.1879 & 1 & -13.9549 & 30.1440 &  2084.0 & 2084.2 & 2097.3 \\
 & (1.6566)& (0.0204)& -& (2.2809) & (0.2597)& &&   \\\hline
KwN  &  8.5874  & 0.1860 & -& -11.2929 & 30.1908 &  2083.1 & 2083.3 & 2096.4 \\
 & (2.0640)& (0.0142)& -& (4.8990) & (0.7101)& &&   \\\hline
EN  &  30694  & 1 & 1 & -526.65 & 146.23 &  2081.8 & 2081.9 & 2091.7 \\
 & (19.5596)& -& -& (32.1783) & (8.1636)& &&   \\\hline
Normal  &  1  & 1& 1 & 76.8762 & 47.3835 &  2136.0 & 2136.1 & 2142.7 \\
 & -& -& -& (3.3339) & (2.3574)& &&   \\\hline
   &  &  &$\lambda$ & $\mu$ & $\sigma$  &  &  & \\\hline
Skew Normal  &    &  & 0.000165 & 76.87 & 47.3835 &  2138.0 & 2138.2 & 2148.0 \\
\hline
\hline \textbf{Skin folds}   & $a$ & $b$ &$c$ & $\mu$ & $\sigma$  &$\rm{AIC}$ &$\rm{CAIC}$&$\rm{BIC}$\\
\hline
McN  &  2.6160  & 0.2046 & 15.9846 & 34.6481 & 20.1861 &  1925.4 & 1925.7 & 1941.9 \\
 & (0.8826)& (0.0252)& (4.3525)& (6.5042) & (1.0988)& &&   \\\hline
BN  &  9.8257  & 0.2330 & 1 & 1.3566 & 24.2344 & 1934.7 & 1934.9 & 1947.9 \\
 & (2.8399)& (0.0179)& -& (4.8932) & (0.8109)& &&   \\\hline
KwN  &  4.8061  & 0.2302 & - & 15.0360 & 22.5331 &  1942.2 & 1942.4 & 1955.4 \\
 & (1.8008)& (0.0189)& -& (6.9872) & (1.4867)& &&   \\\hline
EN  &  16913  & 1 & 1 & -324.71 & 98.7045 &  1935.5 & 1935.6 & 1945.5 \\
 & (15.1522)& -& -& (21.0631) & (5.5416)& &&   \\\hline
Normal  &  1  & 1 & 1 & 69.0218 & 32.4847 &  1983.5 & 1983.6 & 1990.1 \\
 & -& -& -& (2.2856) & (1.6162)& &&   \\\hline
   &  &  &$\lambda$ & $\mu$ & $\sigma$  &  &  & \\\hline
Skew Normal  &    &  & 0.002751 & 68.9505 & 32.4847& 1985.5 & 1985.6 & 1995.4 \\
\hline \textbf{Mass}  & $a$ & $b$ &$c$ & $\mu$ & $\sigma$  &$\rm{AIC}$ &$\rm{CAIC}$&$\rm{BIC}$\\
\hline
McN  &  0.4517  & 0.1729 & 3.3427 & 60.5137 & 5.6781 &  1608.6 & 1608.9 & 1625.1 \\
 & (0.1734)& (0.0192)& (1.1323)& (3.1563) & (0.4095)& &&   \\\hline
BN  &  1.8180  & 0.1639 & 1 & 44.7278 & 7.6276 &  1611.4 & 1611.6 & 1624.6 \\
 & (3.9984)& (0.0408)& - (17.7350) & (2.1213)& &&   \\\hline
KwN  &  1.8071  & 0.1619 & - & 45.8194 & 7.4405 &  1611.3 & 1611.5 & 1624.5 \\
 & (0.6851)& (0.0130)& -& (2.4402) & (0.3717)& &&   \\\hline
EN  &  18.9527  & 1 & 1 & 19.1775 & 24.7761 &  1611.3 & 1611.4 & 1621.2 \\
 & (0.6263)& -& -& (2.3116) & (1.2831)& &&   \\\hline
Normal  &  1  & 1 & 1 & 64.8738 & 13.0378 &  1614.7 & 1614.8 & 1621.3 \\
 & -& -& -& (0.9173) & (0.6487)& &&   \\\hline
  &  &  &$\lambda$ & $\mu$ & $\sigma$  &  &  & \\\hline
Skew Normal  &    &  & 0.000445 & 64.0378 &13.0378&  1616.7 & 1616.8 & 1626.6 \\
\hline
\end{tabular}}
\end{table}

The results are reported in Table~\ref{EMV}. Notice that the three information criteria agree on the model ranking in every case. For the plasma, skin folds, and mass data, the lowest values of the information criteria are obtained from the fit of the McN distribution. Clearly, the McN model having three skewness parameters is preferred
for the three data sets.

\begin{table}
\centering {\caption{LR tests\label{rv}}\vspace*{0.3cm}
\begin{tabular}{c|c|c|c}
\hline \hline \textbf{Plasma} & Hypotheses  & Statistic $w$ & $p$-value  \\
\hline \hline
McN vs BN  & $\mathcal{H}_{0}:c=1$ vs $\mathcal{H}_{1}: \mathcal{H}_{0}\,\mbox{is\,false}$     &  14.6   &  0.00013 \\
McN vs KwN   &$\mathcal{H}_{0}:a=c$ vs $\mathcal{H}_{1}: \mathcal{H}_{0}\,\mbox{is\, false}$      & 13.7    & 0.00021  \\
McN vs EN &$\mathcal{H}_{0}:b=c=1$ vs $\mathcal{H}_{1}: \mathcal{H}_{0}\,\mbox{is\, false}$   &  14.4   & 0.00075   \\
McN vs Normal &$\mathcal{H}_{0}:a=b=c=1$ vs $\mathcal{H}_{1}: \mathcal{H}_{0}\,\mbox{is\, false}$   &  70.6   & $<$0.0001   \\
\hline \hline \textbf{Skin folds} & Hypotheses  & Statistic $w$ & $p$-value  \\
\hline \hline
McN vs BN  & $\mathcal{H}_{0}:c=1$ vs $\mathcal{H}_{1}: \mathcal{H}_{0}\,\mbox{is\,false}$     &  11.3   &  0.00078 \\
McN vs KwN   &$\mathcal{H}_{0}:a=c$ vs $\mathcal{H}_{1}: \mathcal{H}_{0}\,\mbox{is\, false}$      & 18.8    & 0.000015  \\
McN vs EN &$\mathcal{H}_{0}:b=c=1$ vs $\mathcal{H}_{1}: \mathcal{H}_{0}\,\mbox{is\, false}$   &  14.1   & 0.00087   \\
McN vs Normal &$\mathcal{H}_{0}:a=b=c=1$ vs $\mathcal{H}_{1}: \mathcal{H}_{0}\,\mbox{is\, false}$   & 64.1   & $<$0.0001   \\
\hline \hline \textbf{Mass} & Hypotheses  & Statistic $w$ & $p$-value    \\
\hline \hline
McN vs BN  & $\mathcal{H}_{0}:c=1$ vs $\mathcal{H}_{1}: \mathcal{H}_{0}\,\mbox{is\,false}$     &  4.8   &  0.0285 \\
McN vs KwN   &$\mathcal{H}_{0}:a=c$ vs $\mathcal{H}_{1}: \mathcal{H}_{0}\,\mbox{is\, false}$      & 4.7    & 0.0302  \\
McN vs EN &$\mathcal{H}_{0}:b=c=1$ vs $\mathcal{H}_{1}: \mathcal{H}_{0}\,\mbox{is\, false}$   &  0   & 0.0351   \\
McN vs Normal &$\mathcal{H}_{0}:a=b=c=1$ vs $\mathcal{H}_{1}: \mathcal{H}_{0}\,\mbox{is\, false}$   &  12.1   & 0.00705   \\
\hline \hline
\end{tabular}}
\end{table}

A formal test for the need of the third skewness parameter in the McN distribution can be based on the LR statistics. An application of such method to the current data sets furnishes the results shown in Table~\ref{rv}. For the plasma, skin folds and mass data,
we reject the null hypotheses of all
three tests in favor of the McN distribution. The rejection is extremely highly significant for the plasma data, and highly or very highly significant for the skin folds data. This gives clear evidence of the potential need for three skewness parameters when modelling real data.

\begin{figure}
\begin{center}
\includegraphics[width=15cm,height=13cm]{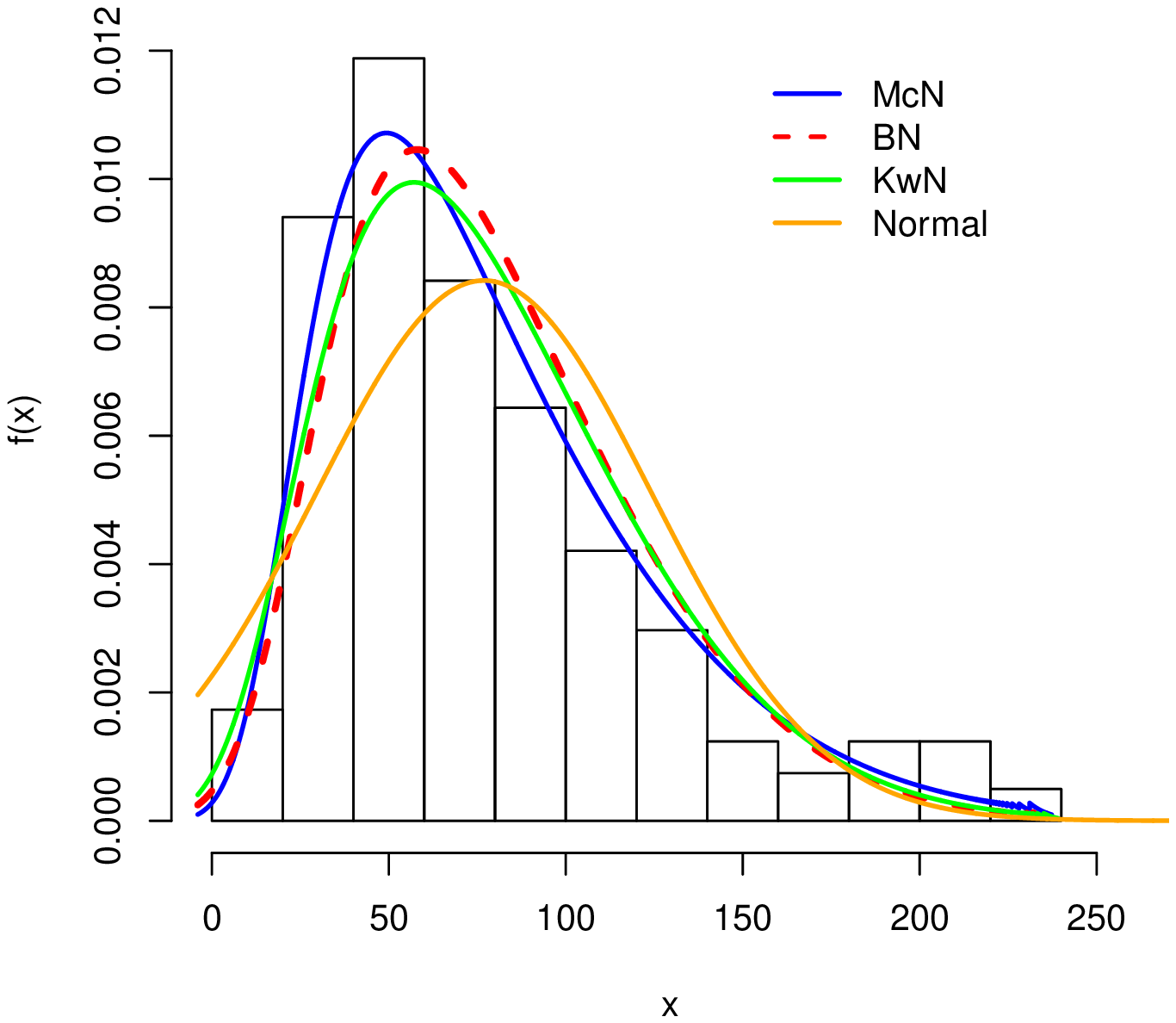}
\caption{Fitted McN, BN, KwN, normal and skew normal densities for the plasma data.}
\label{fe}
\end{center}
\end{figure}

\begin{figure}[htb!]
\begin{center}
\includegraphics[width=15cm,height=13cm]{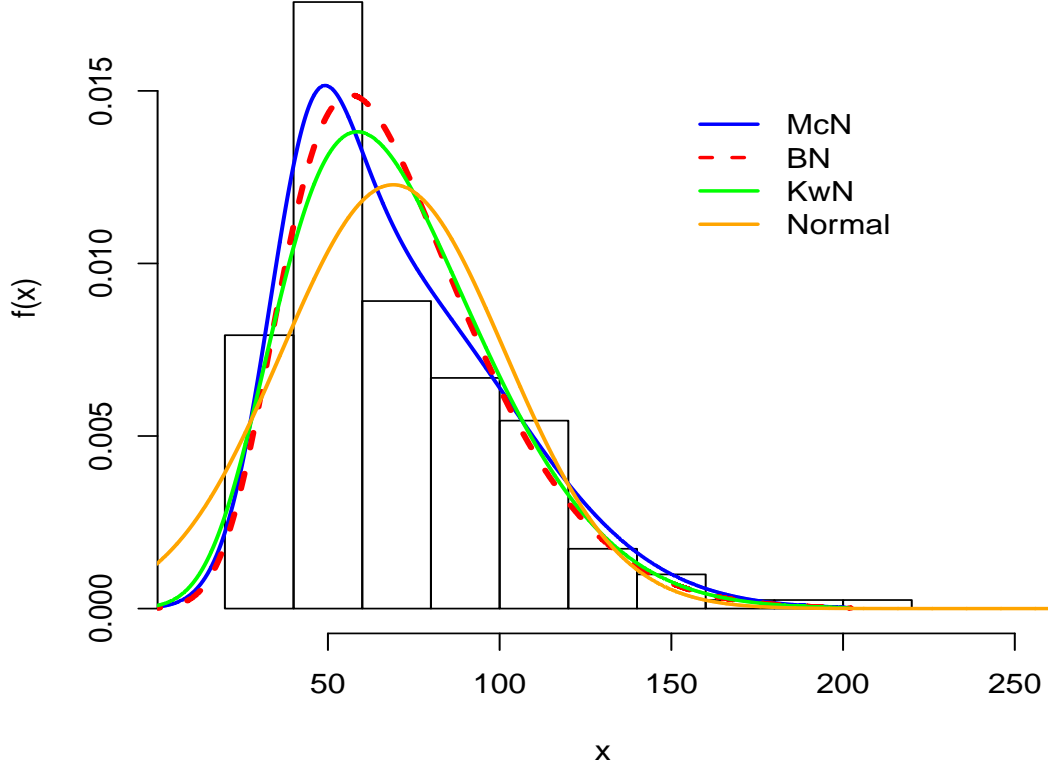}
\caption{Fitted McN, BN, KwN, normal and skew normal densities for the skin folds data.}
\label{ssf}
\end{center}
\end{figure}

\begin{figure}[htb!]
\begin{center}
\includegraphics[width=15cm,height=13cm]{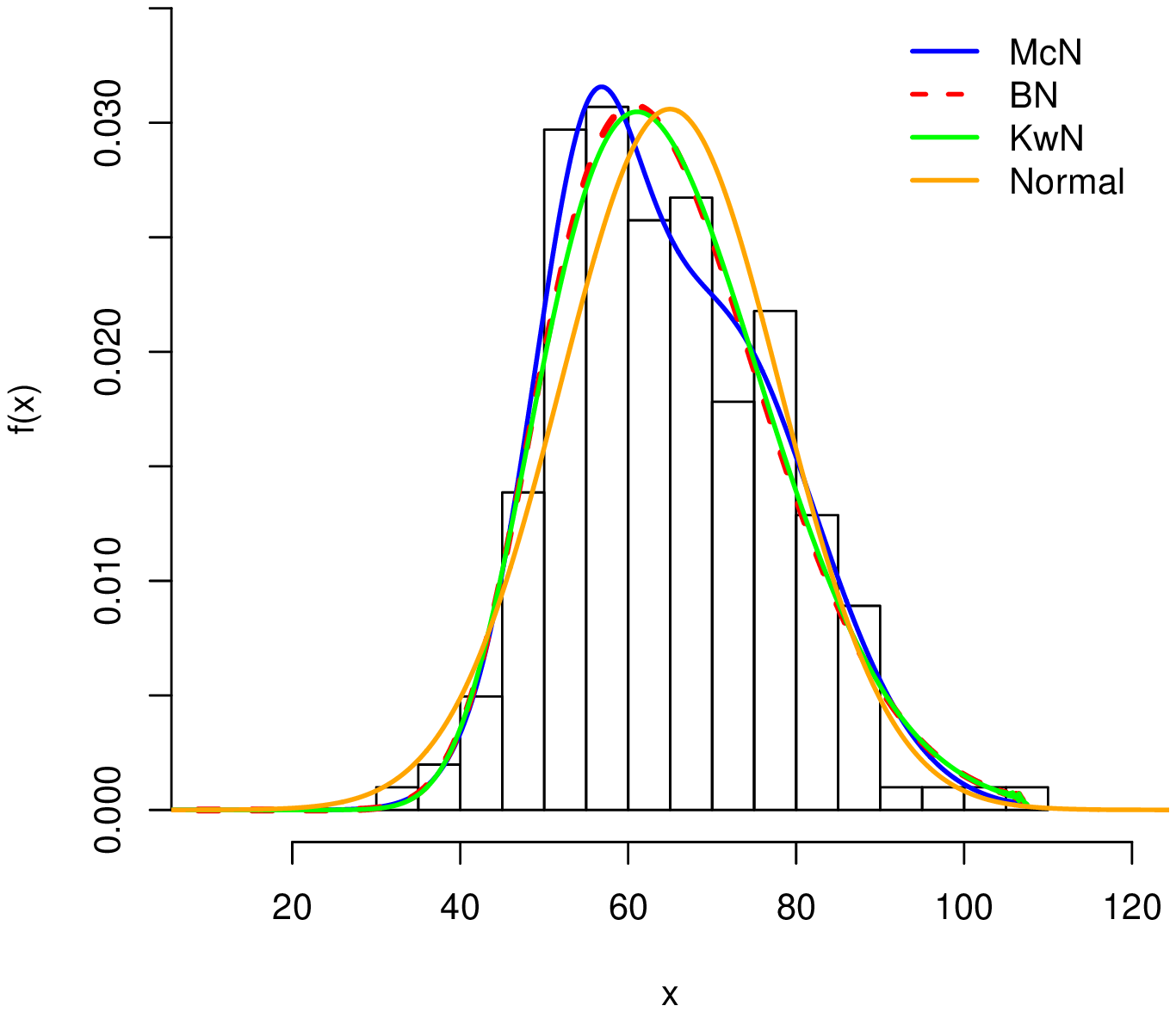}
\caption{Fitted McN, BN, KwN, normal and skew normal densities for the mass data.}
\label{lbm}
\end{center}
\end{figure}

More information is provided by a visual comparison of the histograms of the data with the fitted density functions.
The plots of the fitted McN, BN, KwN, normal and skew normal densities are
given in Figure~\ref{fe} for the plasma data, in Figure~\ref{ssf} for the skin folds data, and in Figure~\ref{lbm} for the mass data. In all these cases,
the McN distribution provides a closer fit to the histograms than the other four sub-models.

\begin{figure}
\begin{center}
\subfigure[]{\includegraphics[width=5.1cm,height=9cm]{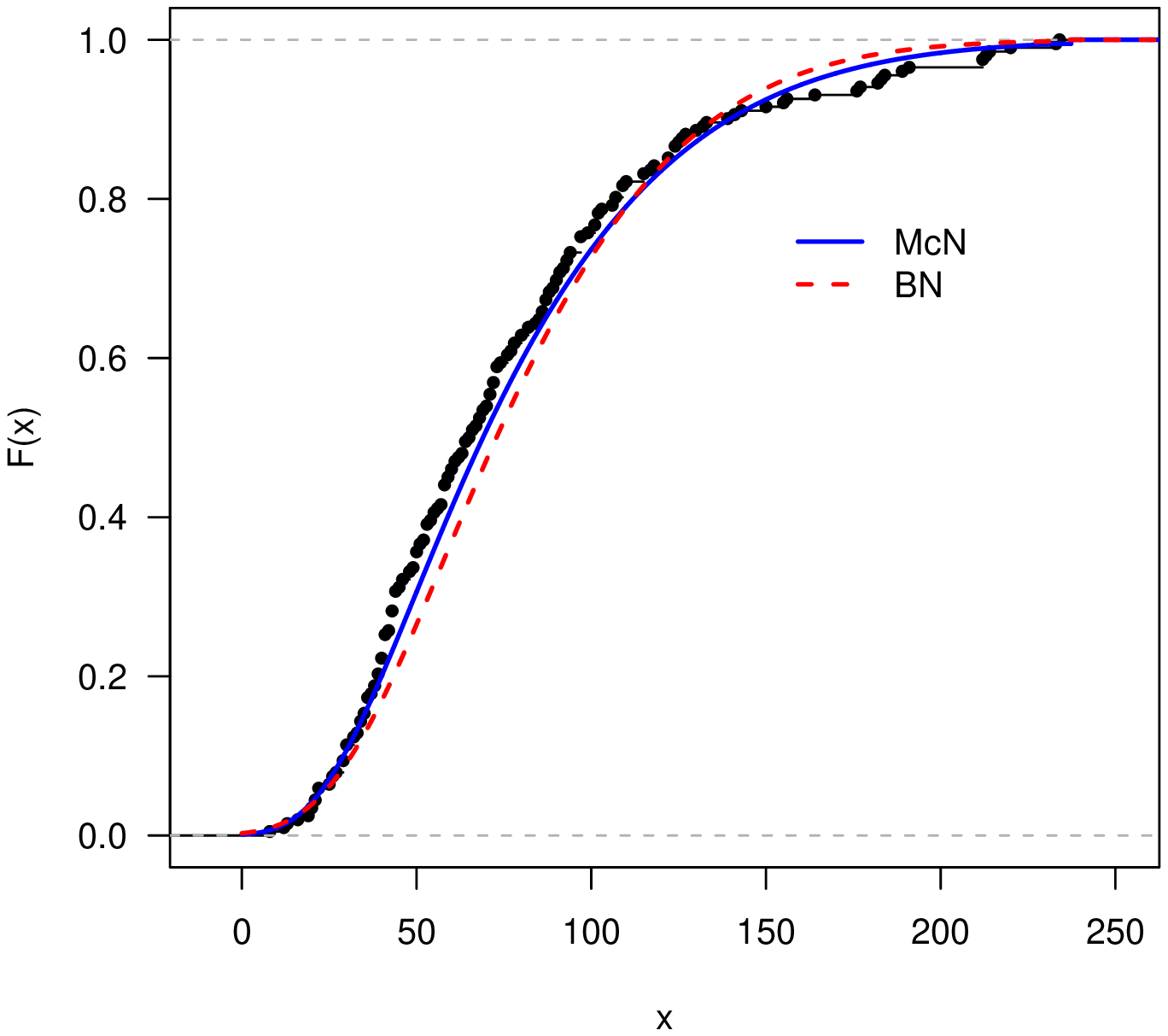}}
\subfigure[]{\includegraphics[width=5.1cm,height=9cm]{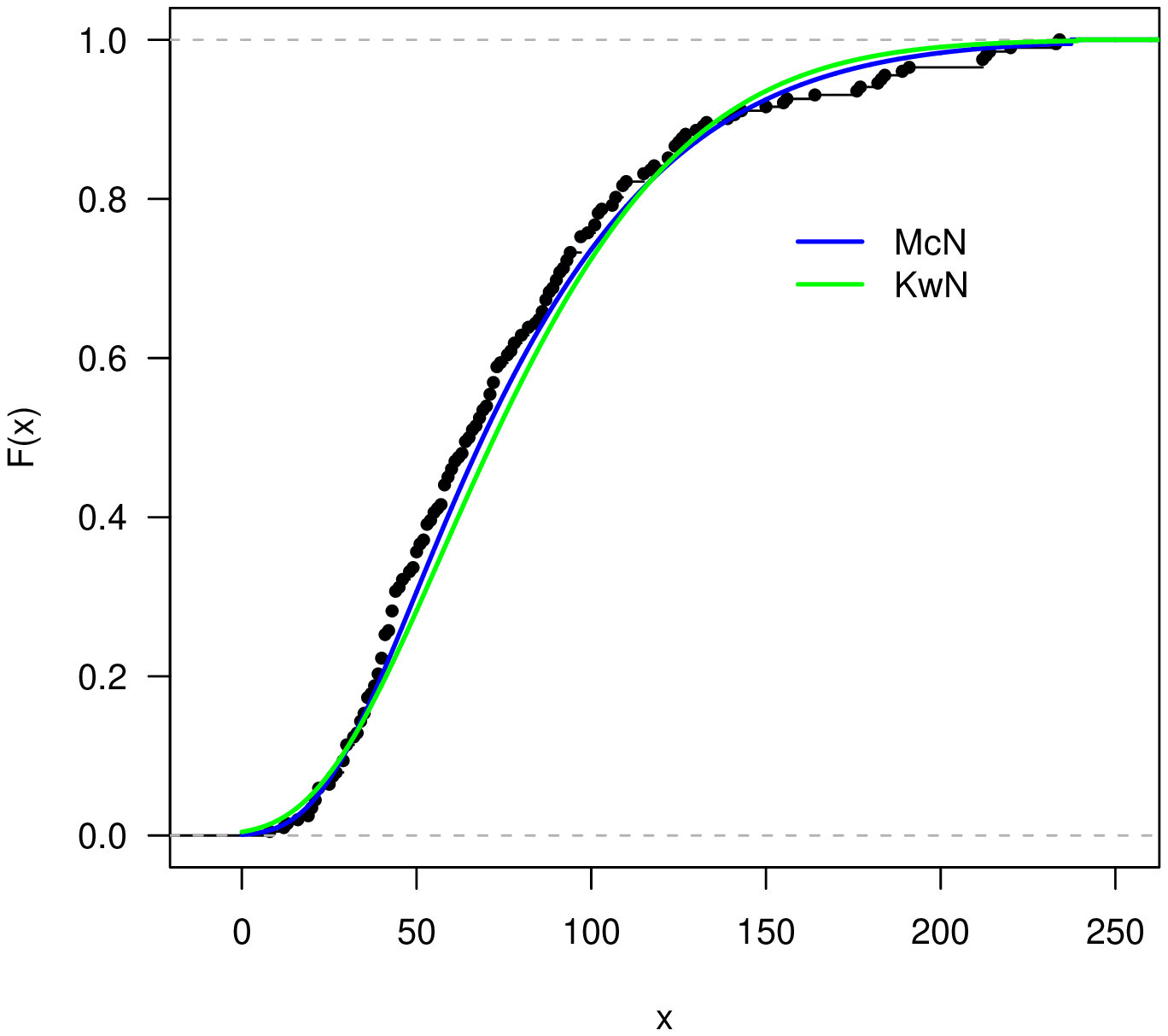}}
\subfigure[]{\includegraphics[width=5.1cm,height=9cm]{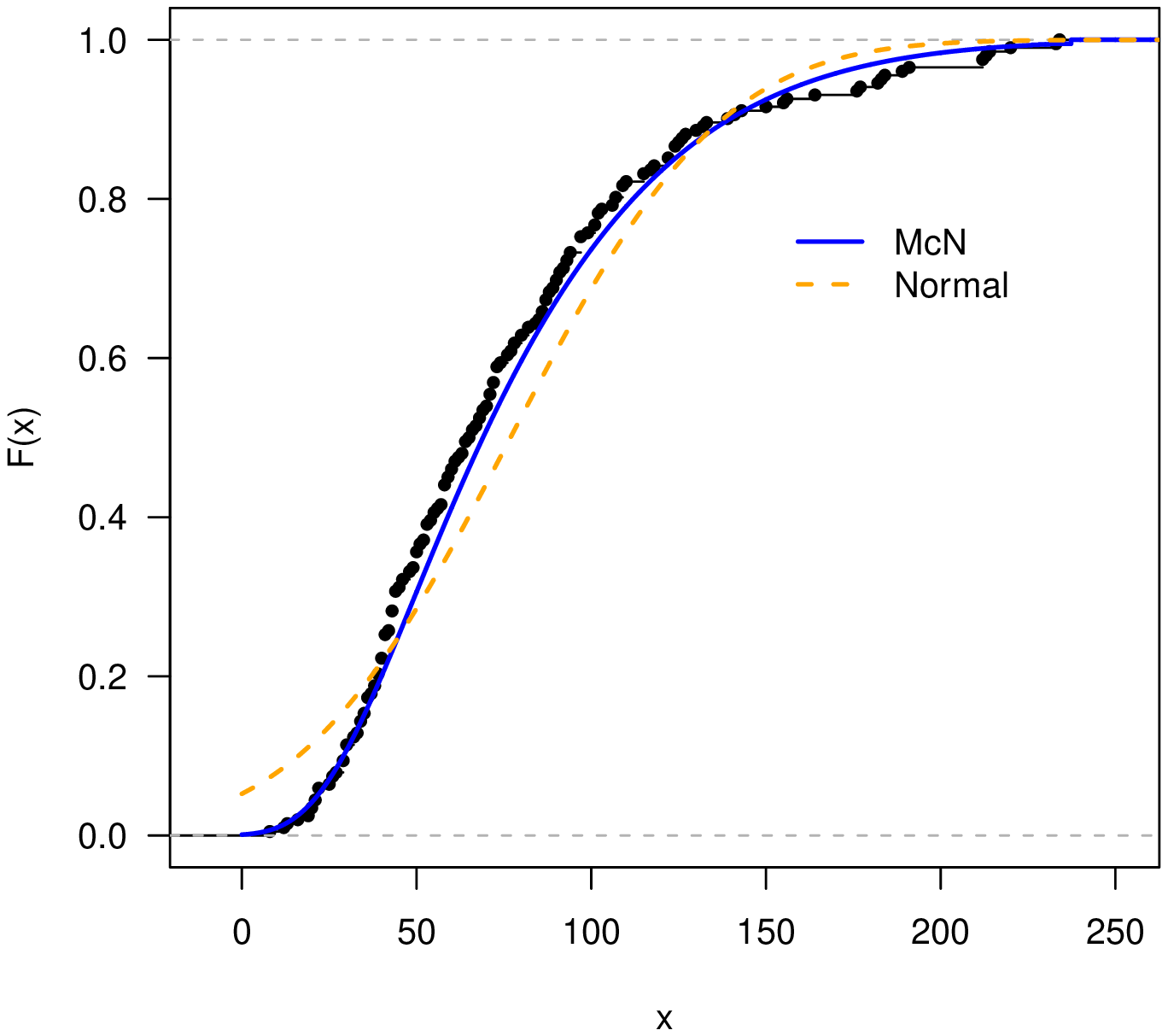}}
\caption{Estimated survival function by fitting the McN distribution and some
other models and the empirical survival for the plasma data.\label{surv1}}
\end{center}
\end{figure}

\begin{figure}
\begin{center}
\subfigure[]{\includegraphics[width=5.1cm,height=9cm]{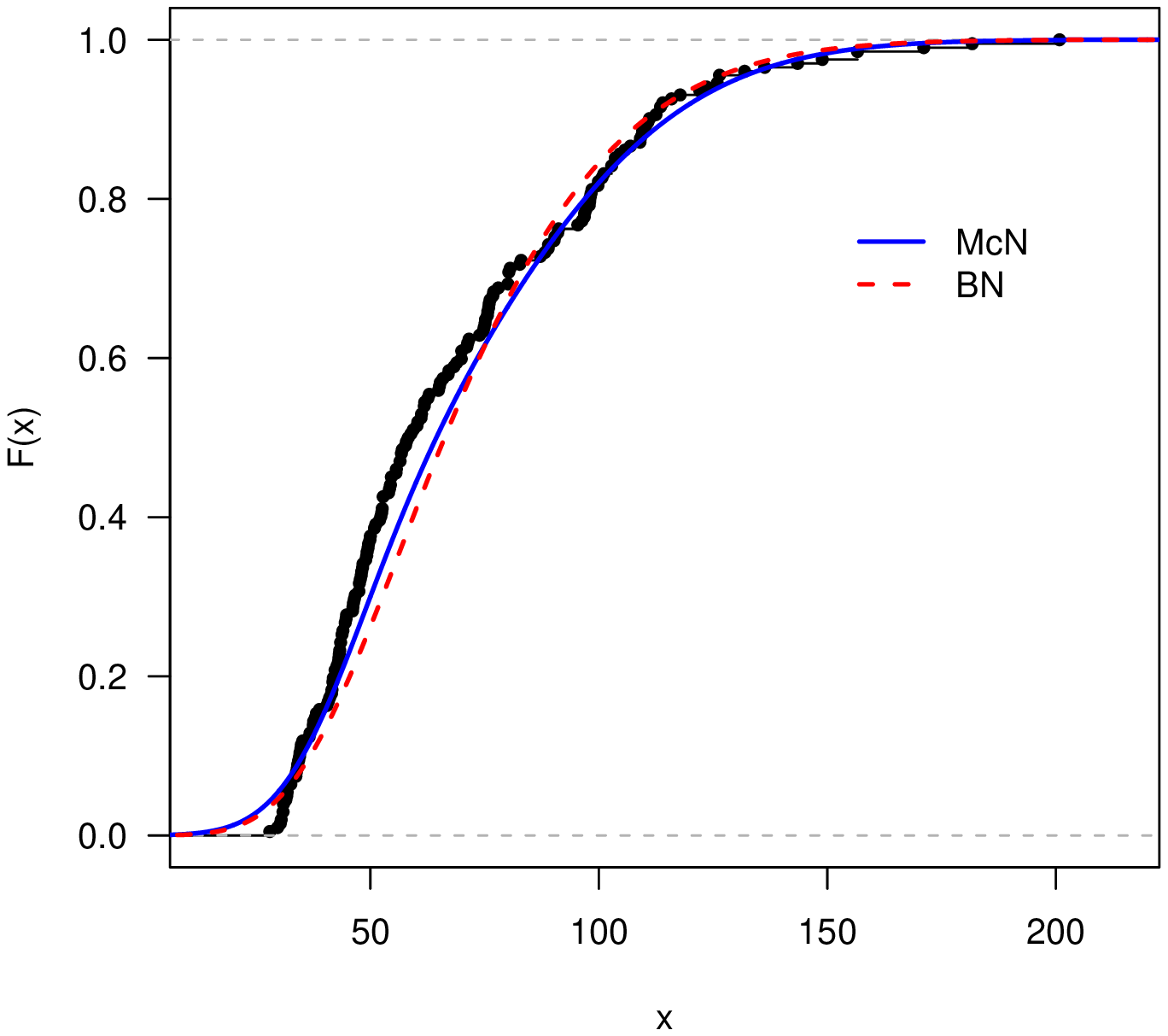}}
\subfigure[]{\includegraphics[width=5.1cm,height=9cm]{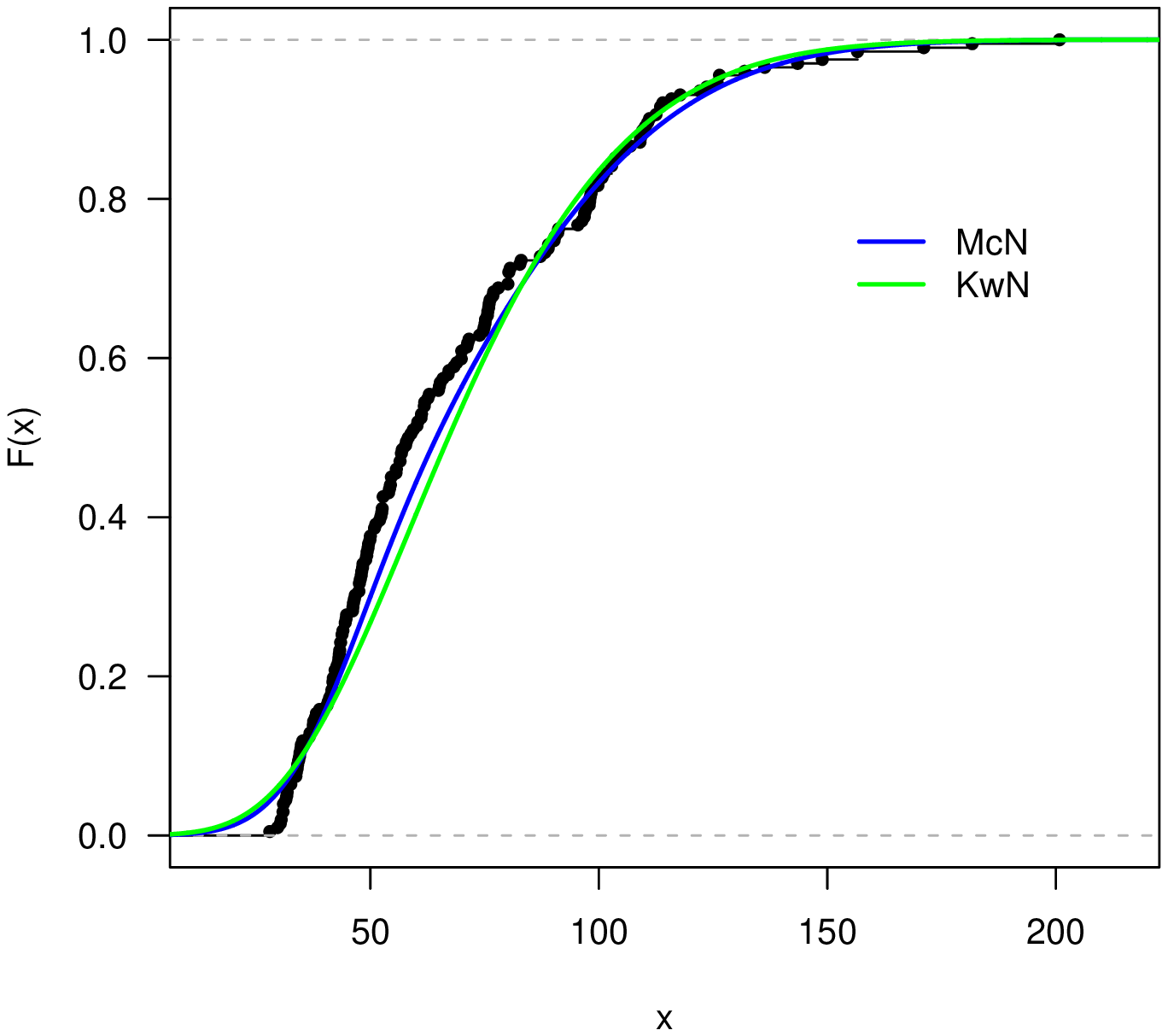}}
\subfigure[]{\includegraphics[width=5.1cm,height=9cm]{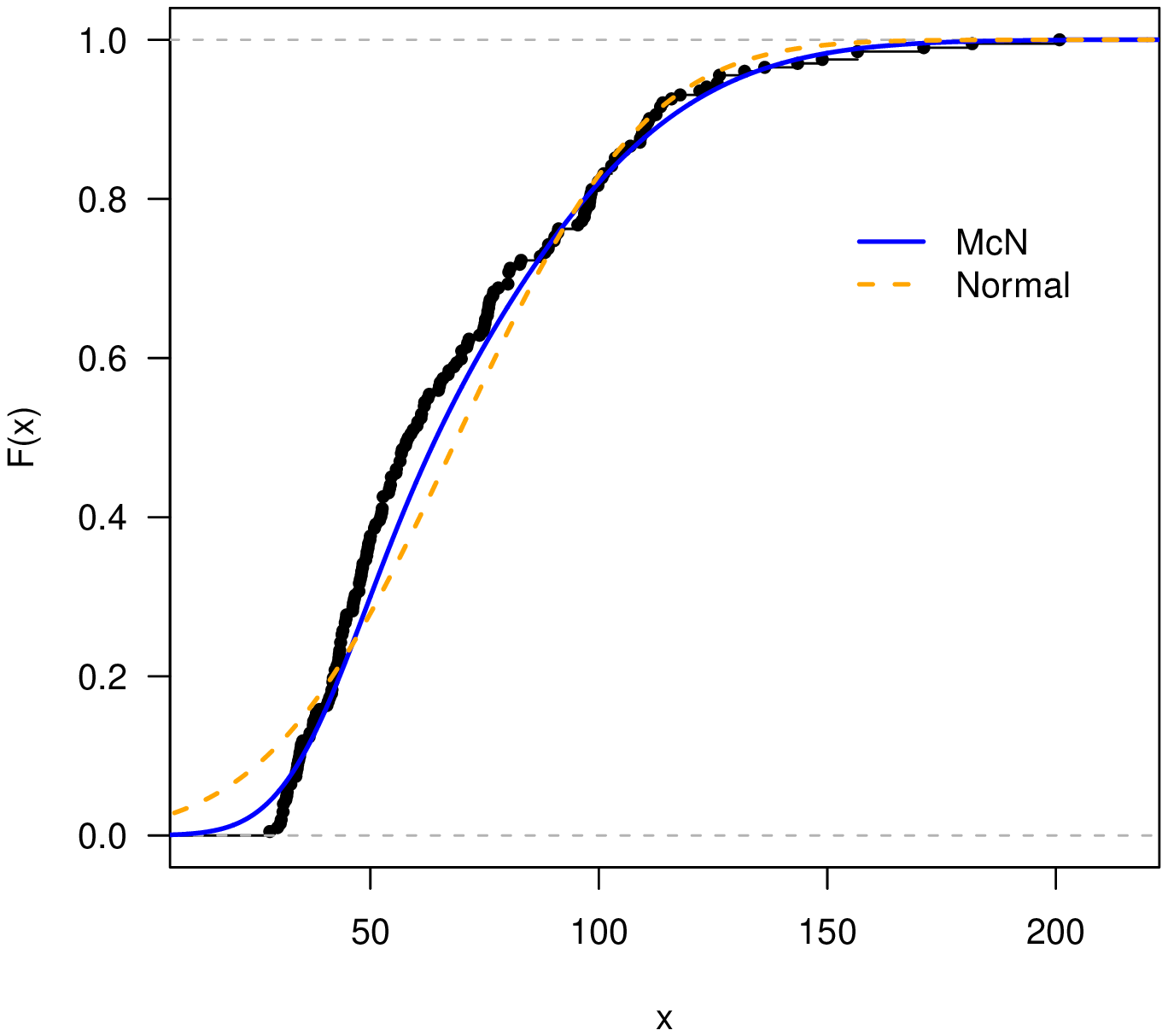}}
\caption{Estimated survival function by fitting the McN distribution and some
other models and the empirical survival for the skin folds data.\label{surv2}}
\end{center}
\end{figure}

\begin{figure}
\begin{center}
\subfigure[]{\includegraphics[width=5.1cm,height=9cm]{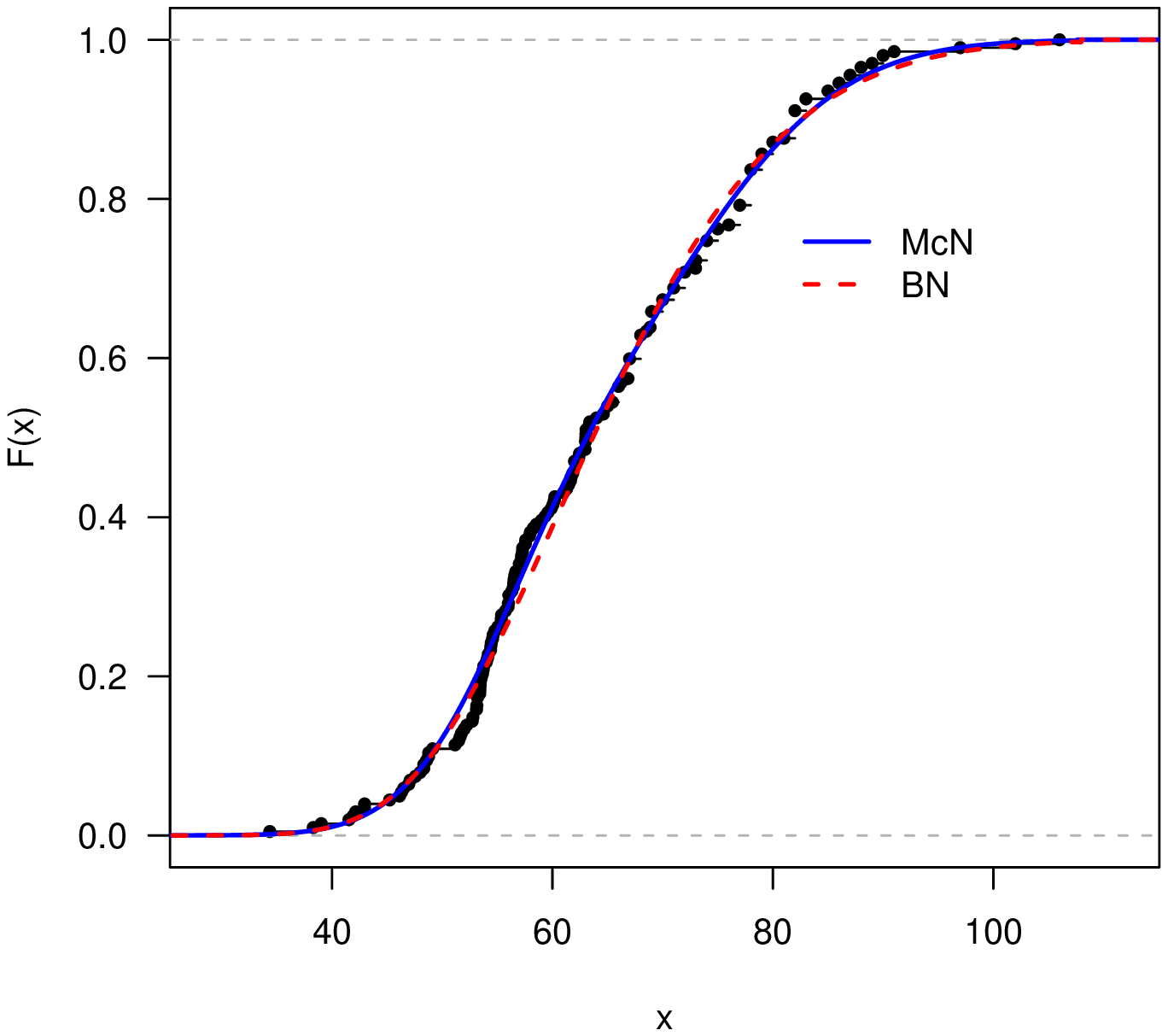}}
\subfigure[]{\includegraphics[width=5.1cm,height=9cm]{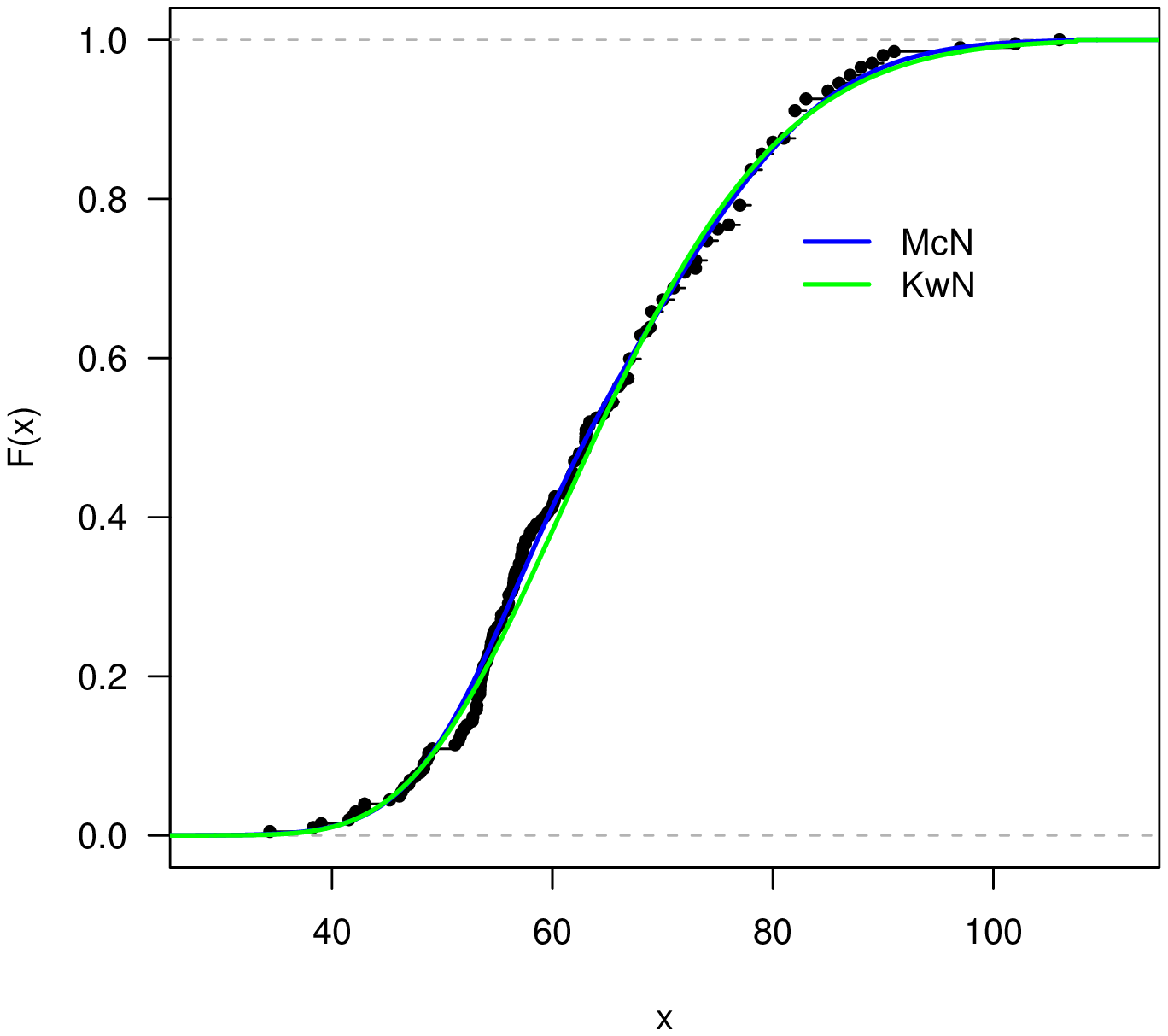}}
\subfigure[]{\includegraphics[width=5.1cm,height=9cm]{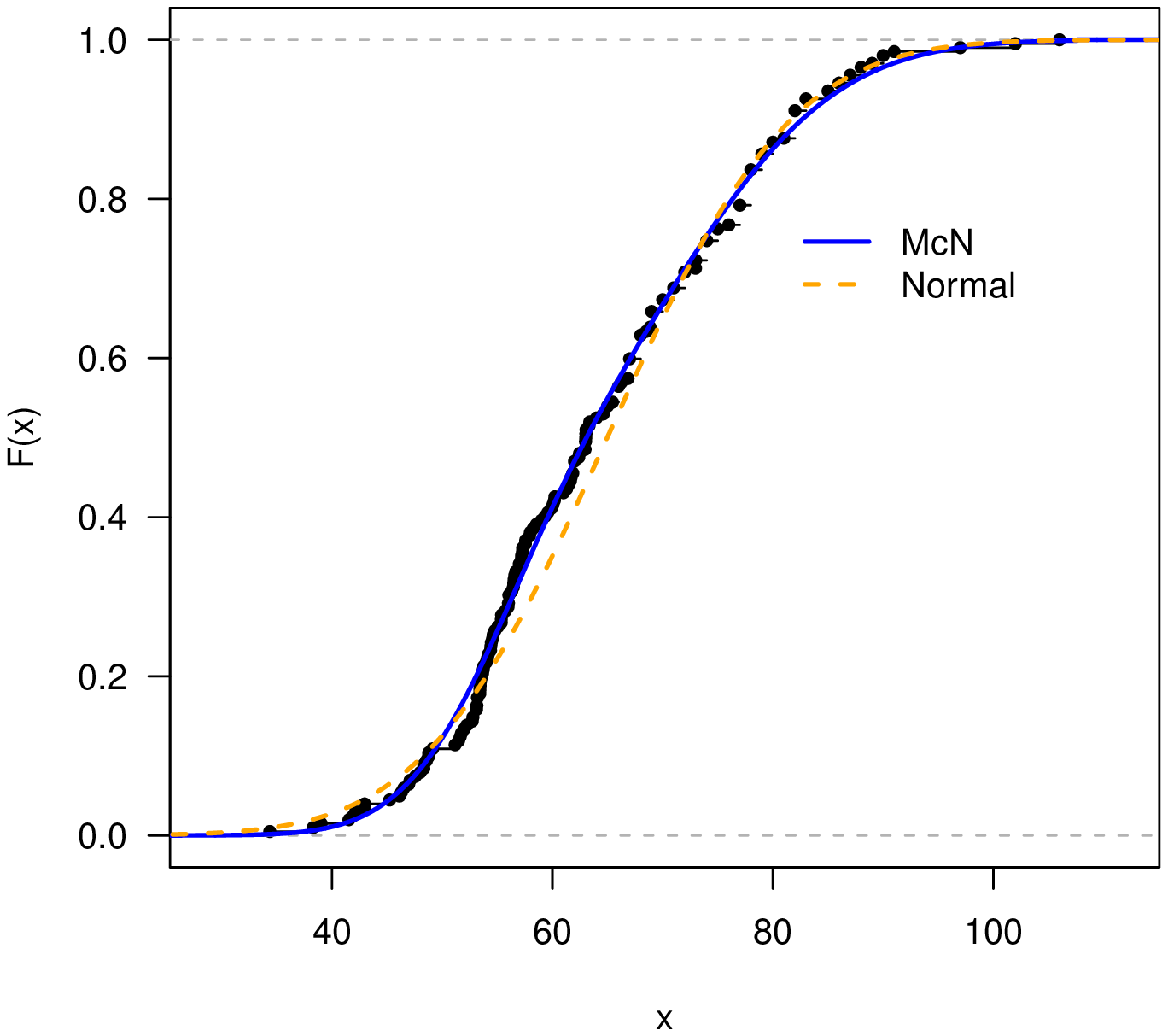}}
\caption{Estimated survival function by fitting the McN distribution and some
other models and the empirical survival for the skin folds data.\label{surv3}}
\end{center}
\end{figure}

In order to assess whether the model is appropriate, the plots of the estimated survival functions
of the McN, BN, KwN and normal distributions and the empirical survival function
are given in Figures~\ref{surv1}, \ref{surv2} and~\ref{surv3} for the plasma, skin folds and mass data, respectively. We conclude that the McN distribution
provides a good fit for the plasma and skin folds data.  For tha mass data, the McN, BN and KwN distributions give reasonable fits, but it is clear that the McN distribution provides a more adequate fit to the histogram of the data.

\section{Conclusions}
\label{section.conclusions}

We present a five parameter lifetime distribution, refereed to as
the McDonald normal (McN) distribution, which includes as
special cases most of the commonly used distributions in the
lifetime literature.
The new distribution is versatile
and analytic tractable for accommodating all four types of density
functions.
Further,
it allows for testing the goodness of fit of several
distributions as submodels.
In fact,
the new model is much more flexible
than the exponentiated normal,
beta normal,
skew normal
and Kumaraswamy normal
models proposed recently.
The McN distribution is capable of improving data fitting substantially
over well-known traditional models. For this new model, powerful para\-me\-tric
methods such as maximum likelihood estimation and
likelihood ratio tests
can be effectively applied to
the analysis of real data sets
to which previous attempts were unsatisfactory and unreliable due to lack of wider
parametric models.
We provide a mathematical treatment of the proposed distribution including
explicit expressions for the ordinary moments, moment generating function and mean deviations,
which hold in generality for any parameter values.
We also give infinite
weighted sums for the moments of the order statistics. Its flexibility, practical
relevance and applicability are demonstrated using three applications
to real data.

\appendix

\section{Elements of $\mathcal{I}(\bm{\theta})$}
\label{section.appendix}

The elements of the observed information matrix $\mathcal{I}(\bm{\theta})$ for the parameters
$(a,b,c,\mu,\sigma)$ are
\begin{eqnarray*}
\mathcal{I}_{a a}(\bm{\theta})=-n[\psi^{(1)}(a)+\psi^{(1)}(a+b)], \qquad \mathcal{I}_{a b}(\bm{\theta})=-n\psi^{(1)}(a+b),\qquad \mathcal{I}_{a c}(\bm{\theta})=\sum\limits_{i=1}^{n}\log[\Phi(z_{i})],
\end{eqnarray*}
\begin{eqnarray*}
\mathcal{I}_{a \mu}(\bm{\theta})=-\frac{c}{\sigma}\sum\limits_{i=1}^{n}\frac{\phi(z_{i})}{\Phi(z_{i})},\qquad
\mathcal{I}_{a \sigma}(\bm{\theta})=-\frac{c}{\sigma}\sum\limits_{i=1}^{n}\frac{z_{i}\phi(z_{i})}{\Phi(z_{i})},\qquad
\mathcal{I}_{b b}(\bm{\theta})=-n[\psi^{(1)}(b)+\psi^{(1)}(a+b)],
\end{eqnarray*}
\begin{eqnarray*}
\mathcal{I}_{b c}(\bm{\theta})=-\sum\limits_{i=1}^{n}\frac{\Phi^{c}(z_{i})\log[\Phi(z_{i})]}{1-\Phi^{c}(z_{i})},
\qquad
\mathcal{I}_{b \mu}(\bm{\theta})
=
\frac{c}{\sigma}\sum\limits_{i=1}^{n}\frac{\phi(z_{i})\Phi^{c-1}(z_{i})}{1-\Phi^{c}(z_{i})},
\qquad
\mathcal{I}_{b \sigma}(\bm{\theta})
=
\frac{c}{\sigma}\sum\limits_{i=1}^{n}\frac{z_{i}\phi(z_{i})\Phi^{c-1}(z_{i})}{1-\Phi^{c}(z_{i})},
\end{eqnarray*}
\begin{eqnarray*}
\mathcal{I}_{c c}(\bm{\theta})
=
-\frac{n}{c^{2}}-(b-1)\sum\limits_{i=1}^{n}\frac{\Phi^{c}(z_{i})\{\log[\Phi(z_{i})]\}^{2}}{[1-\Phi^{c}(z_{i})]^{2}},
\end{eqnarray*}

\begin{eqnarray*}
\mathcal{I}_{c \mu}(\bm{\theta})&=&\frac{a}{\sigma}\sum\limits_{i=1}^{n}\frac{z_{i}\phi(z_{i})}{\Phi(z_{i})}-\frac{b-1}{\sigma}
\sum\limits_{i=1}^{n}\frac{z_{i}\phi(z_{i})\Phi^{c}(z_{i})\{c\log[\Phi(z_{i})]+1\}}{1-\Phi^{c}(z_{i})}
\\&&-\frac{(b-1)c}{\sigma}\sum\limits_{i=1}^{n}
\frac{z_{i}\phi(z_{i})\Phi^{2c-1}(z_{i})\log[\Phi(z_{i})]}{[1-\Phi^{c}(z_{i})]^{2}},
\end{eqnarray*}
\begin{eqnarray*}
\mathcal{I}_{c \sigma}(\bm{\theta})&=&-\frac{a}{\sigma}\sum\limits_{i=1}^{n}\frac{z_{i}\phi(z_{i})}{\Phi(z_{i})}+
\frac{b-1}{\sigma}\sum\limits_{i=1}^{n}\frac{z_{i}\phi(z_{i})\Phi^{c-1}(z_{i})\{c\log[\Phi(z_{i})]+1\}}{1-\Phi^{c}(z_{i})}
\\&&+\frac{(b-1)c}{\sigma}\sum\limits_{i=1}^{n}\frac{z_{i}\phi(z_{i})\Phi^{2c-1}(z_{i})\log[\Phi(z_{i})]}{[1-\Phi^{c}(z_{i})]^{2}},
\end{eqnarray*}
\begin{eqnarray*}
\mathcal{I}_{\mu \mu}(\bm{\theta})&=&-\frac{n}{\sigma^{2}}-\frac{ac-1}{\sigma^{2}}\sum\limits_{i=1}^{n}\frac{\phi(z_{i})[z_{i}\Phi(z_{i})+\phi(z_{i})]}{\Phi^{2}(z_{i})}
\\&&+\frac{(b-1)c}{\sigma^{2}}\sum\limits_{i=1}^{n}\frac{\phi(z_{i})\Phi^{c-1}(z_{i})\{z_{i}-(c-1)\phi(z_{i})\Phi^{-1}(z_{i})\}}{1-\Phi^{c}(z_{i})}
-\frac{(b-1)c^{2}}{\sigma^{2}}\sum\limits_{i=1}^{n}\frac{\phi^{2}(z_{i})\Phi^{2c-2}(z_{i})}{[1-\Phi^{c}(z_{i})]^{2}},
\end{eqnarray*}
\begin{eqnarray*}
\mathcal{I}_{\mu \sigma}(\bm{\theta})&=&-\frac{2}{\sigma^{2}}\sum\limits_{i=1}^{n}z_{i}+\frac{ac-1}{\sigma^{2}}
\sum\limits_{i=1}^{n}\frac{z_{i}\phi(z_{i})[z_{i}\Phi(z_{i})+\phi(z_{i})]}{\Phi^{2}(z_{i})}
+\frac{ac-1}{\sigma^{2}}\sum\limits_{i=1}^{n}\frac{\phi(z_{i})}{\Phi(z_{i})}
\\&&+\frac{(b-1)c}{\sigma^{3}} \sum\limits_{i=1}^{n}\frac{z_{i}^{2}\phi(z_{i})\Phi^{c-1}(z_{i})[1-(c-1)\phi(z_{i})z_{i}^{-1}\Phi^{c-2}(z_{i})]}{1-\Phi^{c}(z_{i})}
\\&&-\frac{(b-1)c^{2}}{\sigma^{3}}\sum\limits_{i=1}^{n}\frac{\phi^{2}(z_{i})\Phi^{2c-2}(z_{i})}{[1-\Phi^{c}(z_{i})]^{2}}
+\frac{(b-1)c}{\sigma^{2}}\sum\limits_{i=1}^{n}\frac{\phi(z_{i})\phi^{c-1}(z_{i})}{1-\Phi^{c}(z_{i})}
\end{eqnarray*}
\begin{eqnarray*}
\mathcal{I}_{\sigma \sigma}(\bm{\theta})&=&-\frac{3}{\sigma^{2}}\sum\limits_{i=1}^{n}z_{i}^{2}-\frac{ac-1}{\sigma^{2}}\sum\limits_{i=1}^{n}
\frac{z_{i}\phi(z_{i})[\Phi(z_{i})(z_{i}^{2}-1)+z_{i}\phi(z_{i})]}{\Phi^{2}(z_{i})}
+\frac{ac-1}{\sigma}\sum\limits_{i=1}^{n}\frac{z_{i}\phi(z_{i})}{\Phi(z_{i})}
\\&&+\frac{(b-1)c}{\sigma^{2}}\sum\limits_{i=1}^{n}\frac{z_{i}\phi(z_{i})\Phi^{c-1}(z_{i})[1-\Phi^{c}(z_{i})]
\{(z_{i}^{2}-1)-(c-1)\Phi^{-1}(z_{i})\}}{[1-\Phi^{c}(z_{i})]^{2}}
\\&&-\frac{(b-1)c}{\sigma^{2}}\sum\limits_{i=1}^{n}
\sum\limits_{i=1}^{n}\frac{z_{i}\phi(z_{i})\Phi^{c-1}(z_{i})}{1-\Phi^{c}(z_{i})},
\end{eqnarray*}
where
$\psi^{(1)}(\cdot)$ is the derivative of the digamma function.

\bibliographystyle{apacite}%
\bibliography{mcn}%

\end{document}